\documentclass{iopart}
\usepackage{iopams}
\usepackage{setstack}
\usepackage{morefloats}
\usepackage{color}
\usepackage{graphicx}
\begin{document}

\title{Relativistic and thermal effects on the magnon spectrum of a ferromagnetic monolayer}

\author{L R\'ozsa$^1$, L Udvardi$^{1,2}$ and L Szunyogh$^{1,2}$}

\address{$^1$ Department of Theoretical Physics, Budapest University of Technology and Economics, Budafoki \'ut 8.~H1111 Budapest, Hungary}
\address{$^2$ Condensed Matter Research Group of Hungarian Academy of Sciences, Budapest University of Technology and Economics, Budafoki \'ut 8.,~H-1111 Budapest, Hungary }

\ead{rozsa@phy.bme.hu}

\begin{abstract}

A spin model including magnetic anisotropy terms and  Dzyaloshinsky-Moriya interactions is studied for the case of a ferromagnetic monolayer with $C_{2v}$ symmetry like Fe/W$(110)$. Using the quasiclassical stochastic Landau-Lifshitz-Gilbert equations, the magnon spectrum of the system is derived using linear response theory. The Dzyaloshinsky-Moriya interaction leads to asymmetry in the spectrum, while the anisotropy terms induce a gap. It is shown that in the presence of lattice defects, both the Dzyaloshinsky-Moriya interactions and the two-site anisotropy lead to a softening of the magnon energies. Two methods are developed to investigate the magnon spectrum at finite temperatures. The theoretical results are compared to atomistic spin dynamics simulations and a good agreement is found between them.

\end{abstract}

\pacs{75.30.Ds, 75.70.Ak, 76.50.+g, 75.40.Gb, 75.30.Gw, 75.40.Mg, 75.10.Hk}

\submitto{\JPCM}

\maketitle

\section{Introduction}

Nanomagnetism has become one of the most intensively studied fields in solid state physics, promising many applications in the near future. Understanding magnetization dynamics in magnetic nanostructures has a key role in the development of magnetic devices. Nanosystems often exhibit different magnetic properties than bulk materials. It is now well established that the magnon spectrum of a thin magnetic film has some properties which may remarkably deviate from the bulk behaviour\cite{Damon,Erickson,Erickson2}. The dipolar coupling between spins influences both the magnon dispersion and the lineshape of the linear response, as studied theoretically\cite{Arias,Arias2,Costa Filho} as well as experimentally\cite{Lindner,Lindner2,Lenz,Xu}.

In case of ultrathin films the magnetic anisotropy and the Dzyaloshinsky-Moriya (DM) interaction\cite{Dzyaloshinsky,Moriya} due to the spin-orbit coupling are more important than the magnetic dipole-dipole interaction. The magnetic anisotropy results in a gap in the magnon spectrum\cite{Prokop,Bergman} necessary to form long-range order in a two-dimensional (2D) system. Recently Zakeri {\it et al.}\cite{Zakeri,Zakeri2012} have detected a so-called magnon Rashba effect\cite{Costa2010} using spin-polarized electron energy loss spectroscopy for a two atomic layer thick Fe film grown on W$(110)$. In the presence of Dzyaloshinsky-Moriya interaction, the energies of the magnons propagating in the [$001$] and [$00\overline{1}$] directions are different. This asymmetry has also been predicted theoretically and calculated from first principles for an Fe monolayer on W(110)\cite{Udvardi}. Recently Cort\`{e}s-Ortu\~{n}o and Landeros\cite{Cortes} studied theoretically the influence of the Dzyaloshinsky-Moriya interactions on thin films using a continuous model. Spin-polarized scanning tunneling microscopy experiments revealed\cite{Bode} that the characteristic length scale of magnetic patterns due to Dzyaloshinsky-Moriya interactions is comparable with the lattice constant, where the application of atomistic models are more relevant than the methods based on the continuous medium model. The asymmetry of magnon energies propagating in opposite directions as a consequence of the Dzyaloshinsky-Moriya interaction was also demonstrated earlier in the bulk system Cs$_{2}$CuCl$_{4}$\cite{Coldea}.

The aim of the present paper is to study the effect of the interplay between different types of interactions and the surface inhomogeneities as well as finite temperature on the linear response of the system using atomistic spin dynamics\cite{Skubic}. The system investigated throughout the paper is a model of an Fe monolayer on W$(110)$. The model contains isotropic and anisotropic exchange between the nearest neighbours, Dzyaloshinsky-Moriya interactions between the next-nearest neighbour pairs and on-site anisotropy with easy axis parallel to the $[001]$ direction. The calculated linear response functions may be comparable to Brillouin light scattering\cite{Hillebrands} and ferromagnetic resonance experiments\cite{Gerhardter}.

The finite temperature phenomena of classical spin systems are usually described by the stochastic Landau-Lifshitz-Gilbert equations\cite{Landau,Gilbert,Brown,Kubo} treated in an atomistic approach\cite{Antropov}. 
The temperature dependence of the exchange stiffness has been calculated in \cite{Atxitia} and, for FePt, an $m^{1.76}$ scaling was found, where
$m$ is the average magnetization. 
In the case of lattice defects, this will lead to a description different from the continuum calculations proposed by Arias {\it et al.}\cite{Arias}. In recent papers\cite{Bergman,Meloche} numerical studies on the temperature dependence of magnetic excitations have been presented, however, the magnon softening as an effect of the dynamical equations was not discussed. There exist several methods for finding approximate analytical solution of the Landau-Lifshitz-Gilbert equation\cite{Lyberatos,Chubykalo,Ma}. The approach proposed by Raikher and Shliomis\cite{Raikher} to describe the linear response of noninteracting spins with on-site anisotropy to external excitations is extended in this paper by including exchange interactions between the spins. A variational approach based on the minimization of the free energy has been introduced to self-consistently renormalize the finite-temperature magnon energies\cite{Bloch,Rastelli}. Although this is a quantumtheoretical treatment of the problem, its classical limit\cite{Gouvea} may be compared to the solution of the Landau-Lifshitz-Gilbert equation. This method is extended here to treat a spin model with Dzyaloshinsky-Moriya interactions. Finally the theoretical calculations will be compared to simulations where the stochastic Landau-Lifshitz-Gilbert equations were solved numerically.

\section{Theory}

\subsection{The model}

The magnetic properties of thin films are often described by a classical Heisenberg model, which turned out to be a very robust and successful scheme. In order to take relativistic effects into account, an extended Heisenberg Hamiltonian is used,
\begin{eqnarray}
H =& \frac{1}{2}\sum_{i \ne j}\boldsymbol{\sigma}_{i}\boldsymbol{J}_{ij}\boldsymbol{\sigma}_{j} + \sum_{i}\big(K_x\sigma_{ix}^{2} + K_y\sigma_{iy}^{2} +K_z\sigma_{iz}^{2}\big) \nonumber
 \\
&- \sum_{i}\boldsymbol{B}_{i}M_{i}\boldsymbol{\sigma}_{i} \, ,  \label{Hamiltonian}
\end{eqnarray}
where $\boldsymbol{\sigma}_{i}$ denotes a unit vector parallel to the average of the magnetization within an atomic sphere at site $i$ and $M_{i}$ stands for the magnitude of the magnetic moment at the given lattice point. In the first term  of the Hamiltonian (\ref{Hamiltonian})  $\boldsymbol{J}_{ij}$ are the $3 \times 3$ exchange interaction matrices, while
the second term represents second order on-site anisotropies, where negative coefficients specify easy magnetization axes. Note that setting $K_{z}$ to zero does not change the Hamiltonian apart from an additive constant. The last term describes the coupling of the spins to an external magnetic field, $\boldsymbol{B}_{i}$. The exchange tensor can be rewritten as
\begin{eqnarray}
 \boldsymbol{J}_{ij}=\frac{1}{3}Tr\big(\boldsymbol{J}_{ij}\big) \cdot \boldsymbol{I} + \bigg[\frac{\boldsymbol{J}_{ij}+\boldsymbol{J}_{ij}^{T}}{2} -\frac{1}{3}Tr\big(\boldsymbol{J}_{ij}\big) \cdot \boldsymbol{I}\bigg] + \frac{\boldsymbol{J}_{ij}-\boldsymbol{J}_{ij}^{T}}{2} \, ,
\end{eqnarray}
where $Tr$ denotes the trace of a matrix, the superscript $T$ labels the transposed matrix and $\boldsymbol{I}$ stands for the unit matrix. The first term corresponds to the isotropic exchange appearing in the usual scalar Heisenberg model. The second term, associated with two-site anisotropy, is a symmetric traceless matrix, similar to the dipolar coupling between the spins. The third term is an antisymmetric matrix, the three independent matrix elements of which are equivalent to the Dzyaloshinsky-Moriya vector, $\boldsymbol{D}_{ij}$ .

\begin{figure}[H]
\centering
\includegraphics[width=15cm,height=12cm]{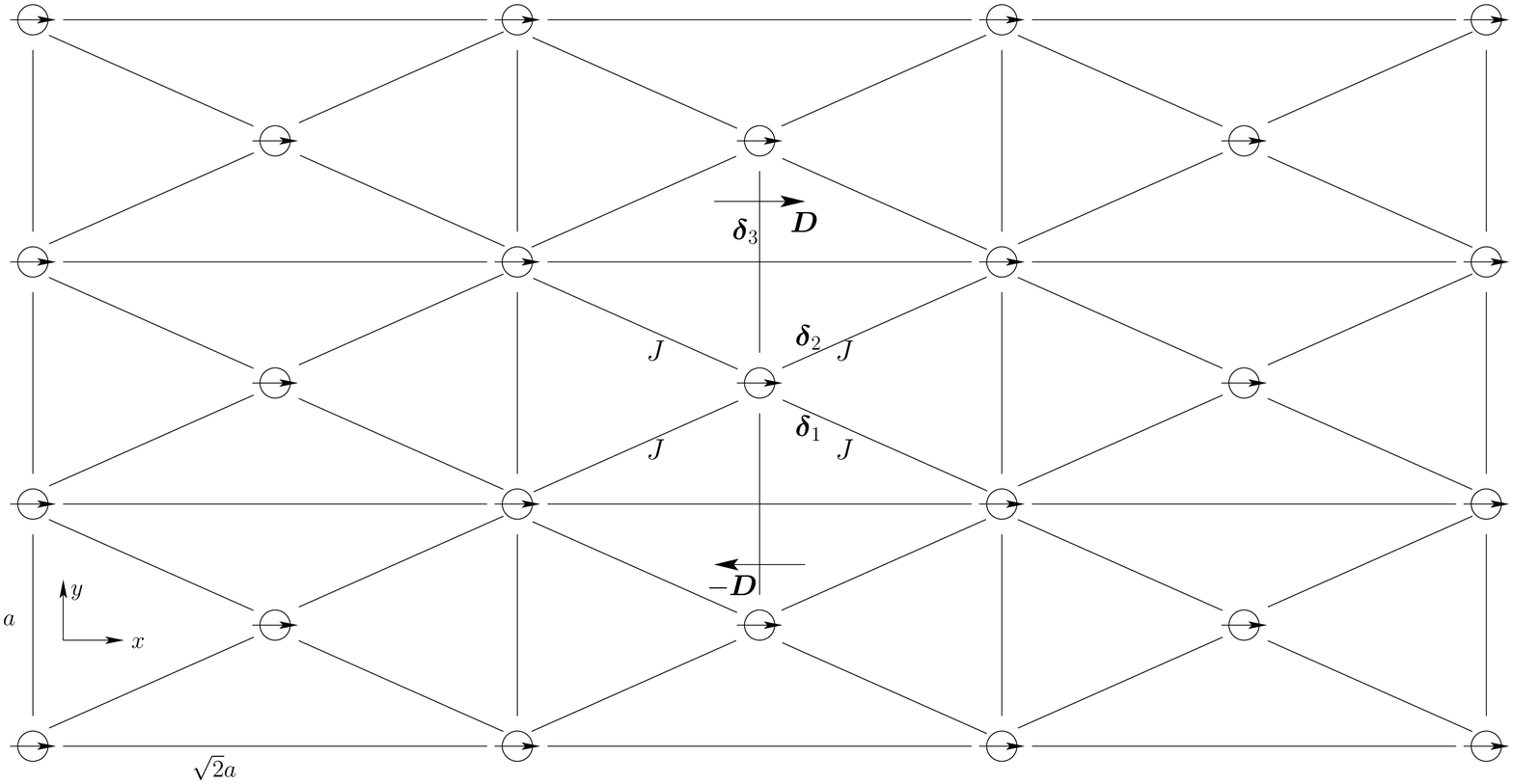}
\caption{Outline of the two-dimensional centered rectangular lattice. The lattice constants along the $x$ and $y$ axes are $\sqrt{2} a$ and $a$, respectively. Small arrows drawn in the circles at the lattice sites correspond to the ground state orientation of the spins. The lattice vectors pointing to nearest neighbour sites are labelled by $\boldsymbol{\delta}_{1}$ and  $\boldsymbol{\delta}_{2}$, while those pointing to a next-nearest neighbour site by $\boldsymbol{\delta}_{3}$.  Displayed are the coupling coefficients appearing in the model of an Fe/W$(110)$ monolayer, $J=J_{xx},J_{yy},J_{zz}$ for nearest neighbours and $D$ for next-nearest neighbours. The orientations of the Dzyaloshinsky-Moriya vectors are also shown by large arrows. 
 \label{szomszed}}
\end{figure}

For our present investigations, a model for an Fe monolayer on the ($110$) surface of bcc W was chosen (see figure \ref{szomszed}). Note that the  $x$, $y$, $z$ axes of the coordinate system are parallel to the $[1\overline{1}0]$, $[001]$ and $[110]$ directions, respectively. Both experimental\cite{Zakeri} and theoretical\cite{Udvardi} studies confirmed that the Dzyaloshinsky-Moriya interaction is present in the system. Only the strongest and most relevant interactions are included in our model that can reproduce the basic properties of the excitation spectrum: nearest-neighbour (NN) exchange interactions, next-nearest-neighbour (NNN) Dzyaloshinsky-Moriya interactions and an easy-axis on-site anisotropy. The exchange tensor contains only diagonal elements $J_{xx}$, $J_{yy}$ and $J_{zz}$ which are slightly different from each other due to the spin-orbit coupling, and $J_{xx}<J_{yy}=J_{zz}<0$ is taken in order to reproduce a ferromagnetic ground state with an easy axis along the $x$ direction. The magnitude of the Dzyaloshinsky-Moriya vector $\boldsymbol{D}_{ij}$ is an order of magnitude smaller than the exchange coupling and parallel to the $[1\overline{1}0]$ direction, constrained to this direction because the $(1\overline{1}0)$ plane containing the next-nearest neighbours is a mirror plane\cite{Dzyaloshinsky}. The third contribution to the energy is the uniaxial anisotropy preferring the $[1\overline{1}0]$ direction, $K_{x}<0$ and $K_{y} \ge 0$.

\subsection{The stochastic Landau-Lifshitz-Gilbert equation}

In an adiabatic approach\cite{Halilov}, the time evolution of the localized magnetic moments in a solid at finite temperature can be described by the stochastic Landau-Lifshitz-Gilbert equations,
\begin{eqnarray}
\frac{\partial \boldsymbol{M}_{i}}{\partial t}=&-\gamma'\boldsymbol{M}_{i}\times(\boldsymbol{B}^{eff}_{i}+\boldsymbol{B}_{i}^{th}) \nonumber
\\
&-\frac{\alpha\gamma'}{M_{i}}\boldsymbol{M}_{i}\times\Big[\boldsymbol{M}_{i}\times(\boldsymbol{B}^{eff}_{i}+
\boldsymbol{B}_{i}^{th})\Big] \, , \label{LLG}
\end{eqnarray}
where $\boldsymbol{M}_{i}=M_{i}\boldsymbol{\sigma}_{i}$ stands for the localized magnetic moment at lattice point $i$, with $M_{i}$ denoting its magnitude, $\alpha$ is the Gilbert damping parameter, and $\gamma'=\gamma/(1+\alpha^{2})$ with the
gyromagnetic ratio, $\gamma=2\mu_{B}/\hbar$. The magnetic field driving the motion of the spins contains two terms.
The deterministic term, $\boldsymbol{B}^{eff}_{i}$, can be obtained from the effective classical Hamiltonian  (\ref{Hamiltonian}),
\begin{eqnarray}
 \boldsymbol{B}^{eff}_{i}=-\frac{\partial H}{\partial \boldsymbol{M}_{i}}
 =-\frac{1}{M_{i}} \frac{\partial H}{\partial \boldsymbol{\sigma}_{i}} \, .
\end{eqnarray}
The thermal term, $\boldsymbol{B}_{i}^{th}$, is proportional to the three-dimensional standard Gaussian white noise, $\boldsymbol{\eta}_{i}$ \cite{Palacios},
\begin{eqnarray}
 \boldsymbol{B}_{i}^{th}(t)=\sqrt{2D_{i}}\boldsymbol{\eta}_{i}(t)\,, \qquad D_{i}=\frac{\alpha}{1+\alpha^{2}}\frac{k_{B}T}{M_{i}\gamma'} \, .
\end{eqnarray}

In the rest of the paper the following simplified notations will be used: $\boldsymbol{B}_{i}^{eff}$ will stand for $\gamma \boldsymbol{B}_{i}^{eff}$, that is it will be measured in frequency dimension. Similarly, the temperature in frequency dimension, $\gamma k_{B}T/M$, will simply be denoted by $T$, where it was used that in case of a monolayer the magnetic moment has the same value at every lattice point, $M_{i}=M$. The thermal field can be written as $ \boldsymbol{B}_{i}^{th}(t)=\Sigma \boldsymbol{\eta}_{i}(t)$ with $\Sigma=\sqrt{2\alpha T}$. The terms $J_{ij}^{\alpha\beta}/M$ and $K_{\alpha}/M$ appearing in the effective field will be replaced with $J_{ij}^{\alpha\beta}$ and $K_{\alpha}$, respectively. Note that we are going to use different model values for these parameters in our calculations. The above definitions and simplifications make it possible to rewrite equation (\ref{LLG}) in the form
\begin{eqnarray}
\frac{\partial \boldsymbol{\sigma}_{i}}{\partial t}=&-\frac{1}{1+\alpha^{2}}\boldsymbol{\sigma}_{i}\times\Big(-\frac{1}{M}\frac{\partial H}{\partial \boldsymbol{\sigma}_{i}}+\Sigma \boldsymbol{\eta}_{i}\Big)\nonumber
\\
& - \frac{\alpha}{1+\alpha^{2}}\boldsymbol{\sigma}_{i}\times\Big[\boldsymbol{\sigma}_{i}\times\Big(-\frac{1}{M}\frac{\partial H}{\partial \boldsymbol{\sigma}_{i}}+\Sigma \boldsymbol{\eta}_{i}\big)\Big]. \label{LLGdless}
\end{eqnarray}

\subsection{The linear response of the system at zero temperature}

\begin{figure}[H]
\centering
\includegraphics[width=7cm,height=5cm]{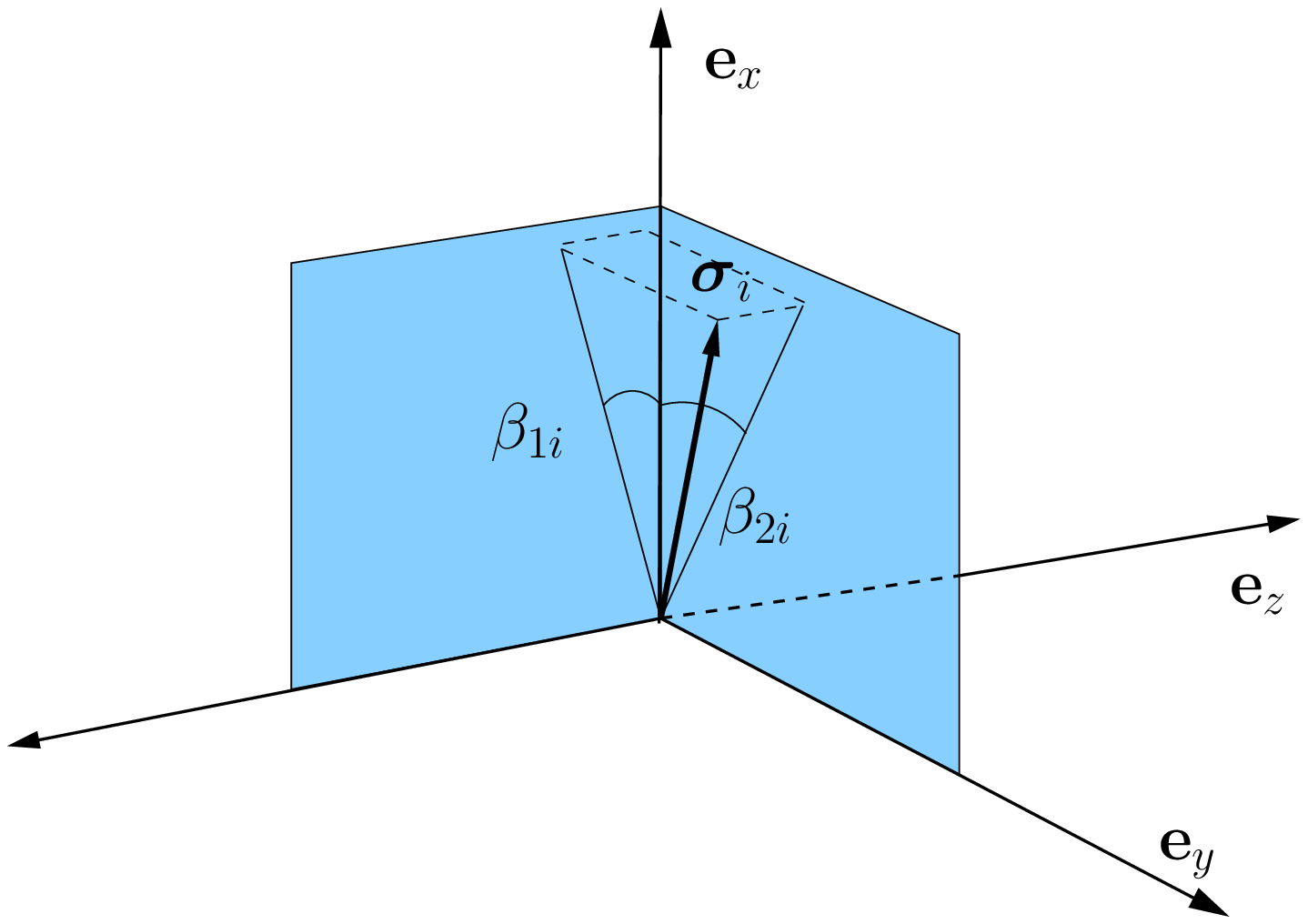}
\caption{Sketch of the axes, $\boldsymbol{e}_{y}$ and $\boldsymbol{e}_{z}$, and the angle variables, $\beta_{1i}$ and $\beta_{2i}$, describing small rotations of a spin around the ground state orientation  $\boldsymbol{e}_{x}$. Also shown are the spin-direction of $\boldsymbol{\sigma}_{i}$ and its projections onto the $xz$ and $xy$ planes.}
\label{szog}
\end{figure}

The set of equations (\ref{LLGdless}) is nonlinear, coupled between the lattice points, and contains multiplicative stochastic noise, all above contributing to the fact that it is quite complicated to find analytic solutions. Firstly the equations shall be solved at zero temperature. The method presented here is a linear approximation, which describes the magnetic excitation of the system close to the ground state and the response to a small dynamic external magnetic field. New variables $\beta_{1i}$ and $\beta_{2i}$ are introduced corresponding to the rotation of the spin vector around the orthogonal vectors $\boldsymbol{e}_{y}$ and $\boldsymbol{e}_{z}$ transverse to the ground state ferromagnetic direction $\boldsymbol{e}_{x}$ as shown in figure \ref{szog}. The $\boldsymbol{\sigma}_{i}$ vector is expanded in the $\beta_{1i}$ and $\beta_{2i}$ variables up to second order as

\begin{eqnarray}
\boldsymbol{\sigma}_{i}=\left[\begin{array}{c}1-\frac{\beta_{1i}^{2}}{2}-\frac{\beta_{2i}^{2}}{2} \\ \beta_{2i} \\ -\beta_{1i}\end{array}\right].
\end{eqnarray}

Using the $C_{2v}$ symmetry of the system and expanding around the ferromagnetic ground state up to second order in the $\beta_{1i}$ and $\beta_{2i}$ variables,  the Hamiltonian (\ref{Hamiltonian}) will take the form
\begin{eqnarray}
H=\sum_{\langle i,j \rangle _{1}}\Bigg[J_{xx}\bigg(1-\frac{\beta_{1i}^2}{2}-\frac{\beta_{2i}^2}{2}\bigg) +  J_{yy}\beta_{2i}\beta_{2j} + J_{zz}\beta_{1i}\beta_{1j}\Bigg] \nonumber
\\
\qquad + \sum_{\langle i,j \rangle _{2}} D^{ij}\bigg(\beta_{1i}\beta_{2j}-\beta_{1j}\beta_{2i}\bigg) + K_x\sum_{i} \bigg(1-\beta_{1i}^{2}-\beta_{2i}^{2}\bigg) \nonumber
\\
\qquad+ K_y\sum_{i} \beta_{2i}^{2} - \sum_{i} \Bigg[B_{ix}\bigg(1-\frac{\beta_{1i}^{2}}{2}-\frac{\beta_{2i}^{2}}{2}\bigg)  + B_{iy}\beta_{2i} - B_{iz}\beta_{1i}\Bigg], \label{quadHam}
\end{eqnarray}
where $\langle i,j \rangle_{1,2}$ denote summations over first and second nearest neighbour pairs, respectively. The Landau-Lifshitz-Gilbert equation can be reformulated using the identity
\begin{eqnarray}
-\frac{\partial H}{\partial \boldsymbol{\sigma}_{i}}\Bigg|_{\bot}&=&-\frac{\partial H}{\partial \beta_{2i}}\boldsymbol{e}_{y} + \frac{\partial H}{\partial \beta_{1i}}\boldsymbol{e}_{z} \, ,
\end{eqnarray}
leading to the zero-temperature expression
\begin{eqnarray}
(1+\alpha^2)\frac{\rmd}{\rmd t}\beta_{2i}=\bigg[-4J_{xx}\beta_{1i} + \sum_{\langle j \rangle _{1}}J_{zz}\beta_{1j} +  \sum_{\langle j \rangle _{2}}D^{ij}\beta_{2j} \nonumber
\\
\qquad - 2K_{x}\beta_{1i} + B_{ix}\beta_{1i} + B_{iz}\bigg] -\alpha \bigg[-4J_{xx}\beta_{2i} + \sum_{\langle j \rangle _{1}}J_{yy}\beta_{2j} \nonumber
\\
\qquad - \sum_{\langle j \rangle _{2}}D^{ij}\beta_{1j} - 2K_{x}\beta_{1i} + 2K_{y}\beta_{2i} + B_{ix}\beta_{2i} - B_{iy}\bigg], \label{eq1}
\end{eqnarray}
\begin{eqnarray}
(1+\alpha^2)\frac{\rmd}{\rmd t}\beta_{1i}=-\bigg[-4J_{xx}\beta_{2i} + \sum_{\langle j \rangle _{1}}J_{yy}\beta_{2j} -  \sum_{\langle j \rangle _{2}}D^{ij}\beta_{1j} \nonumber
\\
\qquad - 2K_{x}\beta_{2i} +2K_{y}\beta_{2i} + B_{ix}\beta_{2i} - B_{iy}\bigg] - \alpha \bigg[-4J_{xx}\beta_{1i} \nonumber
\\
\qquad + \sum_{\langle j \rangle _{1}}J_{zz}\beta_{1j} + \sum_{\langle j \rangle _{2}}D^{ij}\beta_{2j} - 2K_{x}\beta_{1i} + B_{ix}\beta_{1i} + B_{iz}\bigg] \, . \label{eq2}
\end{eqnarray}
Note that according to figure \ref{szomszed}, in the above equations $D^{ij}$ takes the value of $D$ or $-D$.

To uncouple equations (\ref{eq1}) and (\ref{eq2}), we shall use the lattice Fourier transform of the variables and external field,
\begin{eqnarray}
\hat{\beta}_{1(2)}(\boldsymbol{k}_{j})=\frac{1}{\sqrt{n}}\sum_{\boldsymbol{R}_{i}}\rme^{-\rmi\boldsymbol{k}_{j}\cdot\boldsymbol{R}_{i}}\beta_{1i(2i)} \, ,
\\
B_{y(z)}(\boldsymbol{k}_{j})=\frac{1}{\sqrt{n}}\sum_{i} \rme^{-\rmi\boldsymbol{k}_{j}\cdot
\boldsymbol{R}_{i}}B_{iy (iz)} \, ,
\end{eqnarray}
where $\boldsymbol{R}_{i}$ and $\boldsymbol{k}_{j}$ denote real-space lattice vectors and reciprocal-space wave vectors in the first Brillouin zone, respectively, and $n$ is the number of atoms in the 2D lattice.
In the following a small amplitude, time-dependent external excitation $B_{z}(\boldsymbol{k}_{i},t)=B_{z}(\boldsymbol{k}_{i})\cos{\omega t}$ is considered, while, in order to simplify the calculations, $B_{x}(\boldsymbol{k}_{i})=B_{y}(\boldsymbol{k}_{i})=0, J_{yy}=J_{zz}$ and $K_{y}=0$ were chosen. With these assumptions, equations (\ref{eq1}) and (\ref{eq2}) will be reduced to
\begin{eqnarray}
\frac{\rmd}{\rmd t}\hat{\beta}_{2}(\boldsymbol{k}_{i})=\bigg[\hat{J}(\boldsymbol{k}_{i})\hat{\beta}_{1}(\boldsymbol{k}_{i}) +\rmi\hat{D}(\boldsymbol{k}_{i})\hat{\beta}_{2}(\boldsymbol{k}_{i})- \alpha\hat{J}(\boldsymbol{k}_{i}) \hat{\beta}_{2}(\boldsymbol{k}_{i}) \nonumber
\\
\qquad + \rmi\alpha\hat{D}(\boldsymbol{k}_{i}) \hat{\beta}_{1}(\boldsymbol{k}_{i})\bigg] + \frac{B_{z}(\boldsymbol{k}_{i})}{1+\alpha^{2}}\cos{\omega t} \, , \label{lineq1}
\\
\frac{\rmd}{\rmd t}\hat{\beta}_{1}(\boldsymbol{k}_{i})=\bigg[-\hat{J}(\boldsymbol{k}_{i})\hat{\beta}_{2}(\boldsymbol{k}_{i}) +\rmi\hat{D}(\boldsymbol{k}_{i})\hat{\beta}_{1}(\boldsymbol{k}_{i})- \alpha\hat{J}(\boldsymbol{k}_{i}) \hat{\beta}_{1}(\boldsymbol{k}_{i}) \nonumber
\\
\qquad - \rmi\alpha\hat{D}(\boldsymbol{k}_{i}) \hat{\beta}_{2}(\boldsymbol{k}_{i}) \bigg] - \frac{\alpha B_{z}(\boldsymbol{k}_{i})}{1+\alpha^{2}}\cos{\omega t} \, , \label{lineq2}
\end{eqnarray}
where
\begin{eqnarray}
\hat{J}(\boldsymbol{k}_{i})=&\frac{1}{1+\alpha^{2}}\bigg[-4J_{xx}+2J_{yy}\Big(\cos\big(\boldsymbol{k}_{i}\cdot \boldsymbol{\delta}_{1}\big)+\cos\big(\boldsymbol{k}_{i} \cdot \boldsymbol{\delta}_{2}\big)\Big)-2K_{x}\bigg]  \, ,
\\
\hat{D}(\boldsymbol{k}_{i})=&\frac{1}{1+\alpha^{2}}\bigg(2D\sin\big(\boldsymbol{k}_{i} \cdot \boldsymbol{\delta}_{3}\big)\bigg) \, ,
\end{eqnarray}
with the lattice vectors $\boldsymbol{\delta}_{1}$, $\boldsymbol{\delta}_{2}$, and $\boldsymbol{\delta}_{3}$ as depicted in figure \ref{szomszed}.

Introducing the variables $\hat{\beta}_{+}(\boldsymbol{k}_{i})=\hat{\beta}_{2}(\boldsymbol{k}_{i})+\rmi \hat{\beta}_{1}(\boldsymbol{k}_{i})$ and $\hat{\beta}_{-}(\boldsymbol{k}_{i})=\hat{\beta}_{2}(\boldsymbol{k}_{i})-\rmi \hat{\beta}_{1}(\boldsymbol{k}_{i})$, the solution of the differential equations (\ref{eq1}) and (\ref{eq2}) can easily be obtained,
\begin{eqnarray}
\hat{\beta}_{+}(\boldsymbol{k}_{i},t)=&C_{+}\rme^{z_{+}(\boldsymbol{k}_{i})t}+\int_{0}^{t}\frac{1-\rmi \alpha}{1+\alpha^{2}} B_{z}(\boldsymbol{k}_{i})\rme^{z_{+}(\boldsymbol{k}_{i})(t-s)}\cos{\omega s}\rmd s \, ,\label{sol1}
\\
\hat{\beta}_{-}(\boldsymbol{k}_{i},t)=&C_{-}\rme^{z_{-}(\boldsymbol{k}_{i})t}+\int_{0}^{t}\frac{1+\rmi \alpha}{1+\alpha^{2}} B_{z}(\boldsymbol{k}_{i})\rme^{z_{-}(\boldsymbol{k}_{i})(t-s)}\cos{\omega s}\rmd s \, ,\label{sol2}
\end{eqnarray}
where $C_{+}$ and $C_{-}$ are constants, and
\begin{eqnarray}
z_{+}(\boldsymbol{k}_{i})=&(-\alpha-\rmi)\Big[\hat{J}(\boldsymbol{k}_{i})-\hat{D}(\boldsymbol{k}_{i})\Big]=\frac{-\alpha-\rmi}{1+\alpha^{2}}\omega_{0}(\boldsymbol{k}_{i}) \, ,
\\
z_{-}(\boldsymbol{k}_{i})=&(-\alpha+\rmi)\Big[\hat{J}(\boldsymbol{k}_{i})+\hat{D}(\boldsymbol{k}_{i})\Big]=\frac{-\alpha+\rmi}{1+\alpha^{2}}\omega_{0}(-\boldsymbol{k}_{i}) \, ,
\end{eqnarray}
with the characteristic magnon frequencies,
\begin{eqnarray}
\omega_{0}(\boldsymbol{k}_{i})=&-4J_{xx}+4J_{yy}+(-4J_{yy})\Big[1-\cos(\frac{\sqrt{2}}{2}ak_{x})\cos(\frac{1}{2}ak_{y})\Big] \nonumber
\\
&-2D\sin(ak_{y})-2K_{x} \, . \label{spekeq}
\end{eqnarray}

\begin{figure}[H]
\centering
\includegraphics[width=8cm,height=6cm]{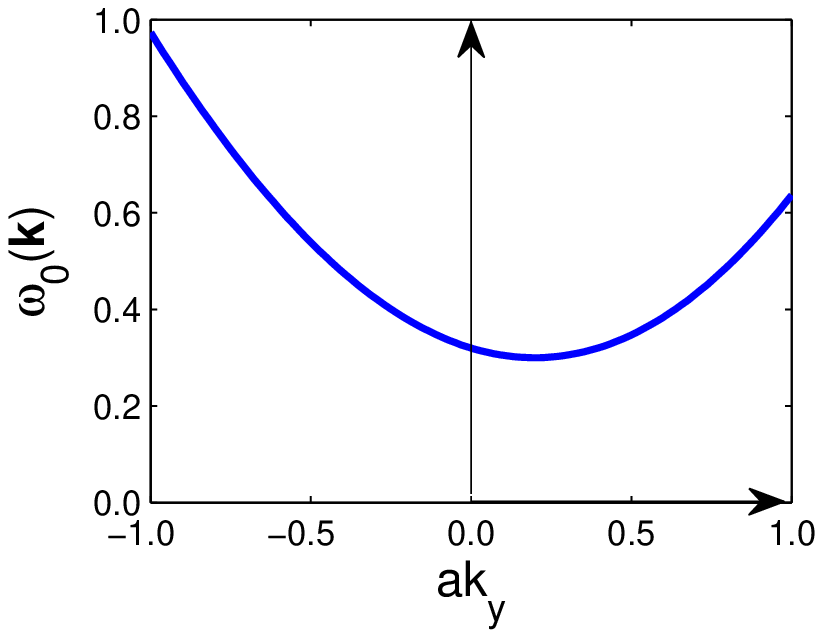}
\caption{Zero temperature magnon spectrum (\ref{spekeq}) with model parameters  $J_{xx}=-1.02, J_{yy}=J_{zz}=-0.99$, $D=0.1$ and $K_{x}=-0.1$. Note the asymmetry of the spectrum with respect to $k_y \rightarrow -k_y$ as a consequence of the Dzyaloshinsky-Moriya interaction.}
\label{spektrum}
\end{figure}

For a typical set of parameters, the spectrum along the $y$ direction is depicted in figure \ref{spektrum}. Two characteristic features of the spectrum should be emphasized. Firstly, the spectrum is not symmetric relative to the $\Gamma$ point in the Brillouin zone, $\omega_{0}(\boldsymbol{k}_{i})\neq\omega_{0}(-\boldsymbol{k}_{i})$, since $\hat{J}(\boldsymbol{k}_{i})=\hat{J}(-\boldsymbol{k}_{i})$, but $\hat{D}(\boldsymbol{k}_{i})=-\hat{D}(-\boldsymbol{k}_{i})$. This has already been demonstrated in previous articles for similar materials, both on theoretical\cite{Udvardi} and experimental\cite{Zakeri} grounds. It is important to
highlight that the Dzyaloshinsky-Moriya interaction has no effect on the magnon energy at the $\Gamma$ point. Secondly, there is a gap in the spectrum due to the on-site and two-site anisotropy terms, the latter one corresponding to the difference between the diagonal elements of the coupling tensor. The value for the gap is $\omega_{0}(\boldsymbol{k}_{i}=\boldsymbol{0})=-4\big(J_{xx}-J_{yy}\big)-2K_{x} > 0$, stabilizing the ferromagnetic ground state. If the magnetic anisotropy is sufficiently small, in the presence of Dzyaloshinsky-Moriya interaction the spectrum may contain excitations with negative energy. In this case, the ground state is usually some sort of chiral state\cite{Vedmedenko, Bode} instead of the ferromagnetic ordering described here. In the long-wavelength limit of the present model, the condition for a spin-spiral ground state reads $|D| > \sqrt{4J_{yy}\big(J_{xx}-J_{yy}\big)+2J_{yy}K_{x} }$.

In the solutions (\ref{sol1}) and (\ref{sol2}), the $C_{+}$ and $C_{-}$ coefficients serve only to fulfill the initial condition of the differential equations, that is the ferromagnetic ground state. The quantities  $z_{+}(\boldsymbol{k}_{i})$ and $z_{-}(\boldsymbol{k}_{i})$ in the exponents have negative real part if the magnon frequencies are positive. In this case, the eigenmodes of the system, that is the first terms of the right-hand sides of (\ref{sol1}) and (\ref{sol2}), decay exponentially and only the response to the external excitation (second terms) survives on a long timescale. As discussed above, negative frequencies indicate that the ferromagnetic state is not stable.

The response of the system to the perturbing field is properly described by the variance of the angle variables defined as
\begin{eqnarray}
S(\boldsymbol{k}_{i},\omega)&=\langle |\hat{\beta}_{+}(\boldsymbol{k}_{i})|^{2} \rangle + \langle |\hat{\beta}_{-}(\boldsymbol{k}_{i})|^{2} \rangle - |\langle \hat{\beta}_{+}(\boldsymbol{k}_{i}) \rangle|^{2} - |\langle \hat{\beta}_{-}(\boldsymbol{k}_{i}) \rangle|^{2} \nonumber
\\
&=2\Big(\langle |\hat{\beta}_{1}(\boldsymbol{k}_{i})|^{2} \rangle + \langle |\hat{\beta}_{2}(\boldsymbol{k}_{i})|^{2} \rangle - |\langle \hat{\beta}_{1}(\boldsymbol{k}_{i}) \rangle|^{2} - |\langle \hat{\beta}_{2}(\boldsymbol{k}_{i}) \rangle|^{2}\Big) \, , \label{S}
\end{eqnarray}
where $\langle \rangle$ simply denotes time averaging for vanishing oscillating terms (eigenmodes).
In this case the expression of $S(\boldsymbol{k}_{i},\omega)$ reads
\begin{eqnarray}
S(\boldsymbol{k}_{i},\omega)=&\frac{\big|B_{z}(\boldsymbol{k}_{i})\big|^{2}}{4(1+\alpha^{2})}\Bigg\{\bigg[\Big(\omega+\frac{\omega_{0}(\boldsymbol{k}_{i})}{1+\alpha^{2}}\Big)^{2}+ \Big(\frac{\alpha\omega_{0}(\boldsymbol{k}_{i})}{1+\alpha^{2}}\Big)^{2}\bigg]^{-1} \nonumber
\\
& +\bigg[\Big(\omega+\frac{\omega_{0}(-\boldsymbol{k}_{i})}{1+\alpha^{2}}\Big)^{2}+ \Big(\frac{\alpha\omega_{0}(-\boldsymbol{k}_{i})}{1+\alpha^{2}}\Big)^{2}\bigg]^{-1} \nonumber
\\
&+\bigg[\Big(\omega-\frac{\omega_{0}(\boldsymbol{k}_{i})}{1+\alpha^{2}}\Big)^{2}+ \Big(\frac{\alpha\omega_{0}(\boldsymbol{k}_{i})}{1+\alpha^{2}}\Big)^{2}\bigg]^{-1} \nonumber
\\
& +\bigg[\Big(\omega-\frac{\omega_{0}(-\boldsymbol{k}_{i})}{1+\alpha^{2}}\Big)^{2}+ \Big(\frac{\alpha\omega_{0}(-\boldsymbol{k}_{i})}{1+\alpha^{2}}\Big)^{2}\bigg]^{-1}\Bigg\} \ , \label{curve}
\end{eqnarray}
which is the sum of four Lorentzian curves. At zero temperature and without damping ($\alpha=0$), the locations of the peaks correspond to the magnon energies at $\boldsymbol{k}_{i}$ and $-\boldsymbol{k}_{i}$ wave vectors. The first two terms have peaks at $\omega<0$ values, because due to the form of the perturbing field the response of the system, $S(\boldsymbol{k}_{i},\omega)$, will be an even function of $\omega$. The other two peaks describe the physical behaviour of the system: if Dzyaloshinsky-Moriya interactions are present, the energies of the $\boldsymbol{k}_{i}$ and $-\boldsymbol{k}_{i}$ magnons will differ, therefore, we will get two peaks instead of a single one. These peaks can be distinguished if the damping is not too large, that is the half-width of the peaks is smaller than the distance between them: $4D\sin\big(\boldsymbol{k}_{i} \cdot \boldsymbol{\delta}_{3}\big)>2\alpha\omega_{0}(\boldsymbol{k}_{i})$. However, at $\boldsymbol{k}_{i}=\boldsymbol{0}$, $S(\boldsymbol{0},\omega)$ will only have a single peak, because the Dzyaloshinsky-Moriya interaction has no effect on the spectrum at the $\Gamma$ point. Similar results for the resonance response of Fe/W($110$) were obtained in \cite{Cortes}, using a macroscopic model of the film. Note that the damping decreases the magnon energies. This effect is, however,
negligible, since for ferromagnetic systems generally $\alpha \ll 1$.

\subsection{Lattice defects\label{lattdef}}

Ferromagnetic resonance is a standard method for studying the linear response to spatially uniform external excitations. In this case, the response of the system will only contain information about excitations with wave vector $\boldsymbol{k}_{i}=\boldsymbol{0}$, the energy of which is, in principle, unaffected by the Dzyaloshinsky-Moriya interactions.  However, if the lattice contains defects, the quasimomentum is not conserved in the system, therefore a spatially uniform external field may create finite wave vector magnons, which are affected by the Dzyaloshinsky-Moriya interactions.

In order to account for a vacancy or a non-magnetic atom replacing a magnetic atom, the value of the spin vector was simply set to zero at the corresponding lattice site. In the model the same simplifications were used as in (\ref{lineq1}) and (\ref{lineq2}), that is $B_{x}(\boldsymbol{k}_{i})=B_{y}(\boldsymbol{k}_{i})=0, J_{yy}=J_{zz}$ and $K_{y}=0$, while a homogeneous external magnetic field was considered, $B_{iz}=B_{z} \cos{\omega t}$. In case of a perfect lattice, equations (\ref{eq1}) and (\ref{eq2}) have clearly the same form for every lattice point, that is the reason why the discrete Fourier transformation could be used to decouple these equations. However if a vacancy is present in the system, these equations take a different form at lattice points neighbouring the defect, since one of the terms will be missing. Apart from these six lattice points, the four nearest-neighbour and the two next-nearest-neighbour sites around the vacancy, the equations again look the same for all the spins. Therefore in this model, only the six neighbours of the vacancy were considered and another ''average'' lattice point, which does not miss a coupling due to the presence of the vacancy. All together one can get a system of coupled equations listed in \ref{lattdefapp} which have to be solved simultaneously. By solving these equations, again the response to the external excitation $S(\boldsymbol{k}_{i}=\boldsymbol{0},\omega)$ was calculated in terms of equation (\ref{S}).



Figures \ref{D0D0J}(a) and \ref{D0D01lyuk50}(a) show the results of numerical calculations for $S(\boldsymbol{k}_{i}=\boldsymbol{0},\omega)$. According to our previous expectations based on the creation of finite wave vector magnons in a disordered system, even in the case of a homogeneous perturbation, if the Dzyaloshinsky-Moriya interactions are present, lattice defects visibly decrease the magnon energy with respect to the case of a perfect lattice. Overall, this is a quite small effect, proportional to the square of the $D/J$ ratio -- this is evidenced by the fact that the direction of the shift does not depend on the sign of this value, that is the same curve is obtained by setting $D/J=\pm 0.1$. On the other hand, the effect is larger if different diagonal exchange coupling coefficients are considered. It was discussed above that, even in a perfect lattice, two-site anisotropy increases the gap in the magnon dispersion at the $\Gamma$ point. As a consequence of the lattice defects, this gap is decreased as well, and this effect is larger than the one due to the Dzyaloshinsky-Moriya interaction. Note that if only on-site anisotropy is taken into account, the lineshape of $S(\boldsymbol{k}_{i}=\boldsymbol{0},\omega)$ is not modified by lattice defects (see appendix \ref{lattdefapp}). A possible explanation for this is that the on-site anisotropy only shifts the magnon spectrum by an additive constant, being independent of the wave vector. In general, it can be concluded that lattice defects may be a source of magnon softening.

\subsection{Finite temperature effects: linear response within a mean field approach\label{sec:mf}}

The stochastic Landau-Lifshitz-Gilbert equations (\ref{LLGdless}) describe the time evolution of the spins at finite temperatures. Linearizing this set of equations is problematic due to the special properties of stochastic calculus. It is still possible to calculate response functions at finite temperatures by solving a system of deterministic differential equations for the first and second moments of the spin components. The method used to treat interacting particles in a mean field approach was originally applied by Raikher and Shliomis\cite{Raikher} for noninteracting magnetic particles possessing an easy orientation axis. Here it is generalized to interacting particles in a mean field approach. The dynamical equations (\ref{LLGdless}) can be rewritten in Cartesian indices as
\begin{eqnarray}
(1+\alpha^{2})\rmd\sigma_{i\alpha}=-\varepsilon_{\alpha\beta\gamma}\sigma_{i\beta}B_{i\gamma}^{eff}\rmd t + \alpha B_{i\alpha}^{eff}\rmd t-\alpha\sigma_{i\alpha}\sigma_{i\beta}B_{i\beta}^{eff}\rmd t \nonumber
\\
\qquad- \Sigma^{2}\sigma_{i\alpha}\rmd t -\varepsilon_{\alpha\beta\gamma}\Sigma\sigma_{i\beta}\rmd W_{i\gamma} + \alpha \Sigma\rmd W_{i\alpha}- \alpha\Sigma\sigma_{i\alpha}\sigma_{i\beta}\rmd W_{i\beta} \ ,
\end{eqnarray}
which is more common in stochastic calculus. Note that in the above expressions a sum has to be taken over the Cartesian indices occurring twice. $\rmd W_{i\alpha}$ stands for the differential form of the one-dimensional Wiener process, with the usual properties: an almost surely continuous Gaussian stochastic process starting from $W_{i\alpha}(0)=0$ with first and second moments $\langle W_{i\alpha}(t) \rangle=0$, $\langle W_{i\alpha}(t)W_{j\beta}(t') \rangle=\delta_{ij} \delta_{\alpha\beta} \rm{min}\{t,t'\}$, respectively. Remember that equations (\ref{LLGdless}) must be interpreted in the Stratonovich sense of stochastic calculus to yield the correct thermal equilibrium properties\cite{Palacios}. In addition, here the equivalent It\^o form\cite{Kloeden} of the equation was used, hence the extra term $- \Sigma^{2}\sigma_{i\alpha}\rmd t$ , which does not appear when simply calculating the vectorial products. It is straightforward to calculate the equations for the first and second moments, in the latter, case using It\^o's formula,
\begin{eqnarray}
(1+\alpha^{2})\frac{\rmd}{\rmd t}\langle\sigma_{i\alpha}\rangle=-\varepsilon_{\alpha\beta\gamma}\langle\sigma_{i\beta}B_{i\gamma}^{eff}\rangle + \alpha \langle B_{i\alpha}^{eff}\rangle-\alpha\langle\sigma_{i\alpha}\sigma_{i\beta}B_{i\beta}^{eff}\rangle \nonumber
\\
\qquad - \Sigma^{2}\langle\sigma_{i\alpha}\rangle \, ,  \label{dmom1/dt} \\
(1+\alpha^{2})\frac{\rmd}{\rmd t}\langle\sigma_{i\alpha}\sigma_{m\beta}\rangle = -\varepsilon_{\alpha\gamma\delta}\langle\sigma_{i\gamma}B_{i\delta}^{eff}\sigma_{m\beta}\rangle - \varepsilon_{\beta\gamma\delta}\langle\sigma_{m\gamma}B_{m\delta}^{eff}\sigma_{i\alpha}\rangle \nonumber
\\
\qquad + \alpha \langle B_{i\alpha}^{eff}\sigma_{m\beta}\rangle + \alpha \langle B_{m\beta}^{eff}\sigma_{i\alpha}\rangle - \alpha\langle\sigma_{i\alpha}\sigma_{m\beta}\sigma_{i\gamma}B_{i\gamma}^{eff}\rangle \nonumber
\\
\qquad - \alpha\langle\sigma_{i\alpha}\sigma_{m\beta}\sigma_{m\gamma}B_{m\gamma}^{eff}\rangle - 2\Sigma^{2}\langle\sigma_{i\alpha}\sigma_{m\beta}\rangle \nonumber
\\
\qquad + \delta_{im}\Sigma^{2}(\delta_{\alpha\beta}-\langle\sigma_{i\alpha}\sigma_{i\beta}\rangle) \, ,
\label{dmom2/dt}
\end{eqnarray}
where $\langle\rangle$ denotes stochastic expectation value. In order to calculate the linear response, the effective field is divided into an unperturbed part and a perturbation, $\boldsymbol{B}_{i}^{eff}=\boldsymbol{B}_{i}^{eff,0}+\boldsymbol{B}_{i}^{eff,pert}(t)$, and the perturbation will be a time-dependent external magnetic field as in the previous section. In absence of the perturbing field, the equilibrium distribution of the spins corresponds to the Boltzmann distribution, as this is the property that determines the standard deviation of the stochastic noise at finite temperatures (see \cite{Palacios}). The probability density function is then given by
\begin{eqnarray}
P_0(\{\boldsymbol{\sigma}_{i}\})=\frac{1}{Z_0}{\rm e}^{-\frac{1}{T} H_{0}(\{\boldsymbol{\sigma}_{i}\})} \, ,
\end{eqnarray}
where $H_0(\{\boldsymbol{\sigma}_{i}\})$ is the unperturbed Hamiltonian and $Z_0$ is the corresponding partition function. If the system is in equilibrium, the moments of the spins will be labelled by subscript $0$. If the perturbation is present, the time-dependent probability density function can be approximated as\cite{Raikher}
\begin{eqnarray}
P^{pert}(\{\boldsymbol{\sigma}_{i}\})=\big(1+(\sigma_{k\varepsilon}-\langle\sigma_{k\varepsilon}\rangle_{0}) a_{k\varepsilon}(t)+(\sigma_{k\varepsilon}\sigma_{l\zeta}-\langle\sigma_{k\varepsilon}\sigma_{l\zeta}\rangle_{0}) \nonumber
\\
\qquad\times b_{k\varepsilon,l\zeta}(t)\big)P_0(\{\boldsymbol{\sigma}_{i}\}) \, , \label{Ppert}
\end{eqnarray}
where the time-dependence only appears in the $a_{k\varepsilon}(t)$ and $b_{k\varepsilon,l\zeta}(t)$ quantities, which are supposed to be linear in the perturbing field. Using this assumption, one can rewrite equations (\ref{dmom1/dt}) and (\ref{dmom2/dt}), which turn into a system of linear differential equations for the $a_{k\varepsilon}(t)$ and  $b_{k\varepsilon,l\zeta}(t)$ functions.

To simplify the calculations, suppose that the unperturbed Hamiltonian has the form,
\begin{eqnarray}
H_{0}(\{\boldsymbol{\sigma}_{i}\})=\sum_{\langle i,j \rangle_{1}}\Big(J_{xx} \sigma_{ix}\sigma_{jx}+J_{yy} \sigma_{iy}\sigma_{jy}+J_{zz} \sigma_{iz}\sigma_{jz}\Big) \nonumber
\\
\qquad+ K_{x}\sum_{i}\sigma_{ix}^{2} \, , \label{H0}
\end{eqnarray}
that is we only consider diagonal coupling coefficients between the nearest neighbour spins and easy-axis anisotropy. This system is time-reversal invariant, therefore all expectation values will vanish which contain an odd number of spin components. Furthermore, it is easy to see that the energy of the system does not change if we replace $\sigma_{ix}$ with $-\sigma_{ix}$ at all lattice points, and the same holds for the $y$ and $z$ components. This leads to the property that only such expectation values are different from zero, which contain an even number of spin components separately for the $x,y,z$ directions. Note that including an external magnetic field in the Hamiltonian would break time-reversal invariance, while including Dzyaloshinsky-Moriya interaction would break the latter symmetry, making the calculations more complicated. As before, $J_{yy}=J_{zz}$ was assumed and a perturbing field pointing towards the $z$ axis, $B_{iz}$. In this case, only the coefficients $a_{kz}(t)$ and $b_{kx,ly}(t)$ differ from zero and they are determined by the following coupled equations,
\begin{eqnarray}
(1+\alpha^{2})\langle\sigma_{iz}\sigma_{kz}\rangle_{0}\frac{\rmd}{\rmd t}a_{kz}= \langle B_{ix}^{eff,0}\sigma_{iy}\sigma_{kx}\sigma_{ly}\rangle_{0}2b_{kx,ly}
\label{eqa} \\
\quad-\langle B_{iy}^{eff,0}\sigma_{ix}\sigma_{kx}\sigma_{ly}\rangle_{0}2b_{kx,ly}+\alpha\langle B_{iz}^{eff,0}\sigma_{kz}\rangle_{0}a_{kz} \nonumber
\nonumber \\
\quad-\alpha\langle B_{i\alpha}^{eff,0}\sigma_{i\alpha}\sigma_{iz}\sigma_{kz}\rangle_{0}a_{kz} -\Sigma^{2}\langle\sigma_{iz}\sigma_{kz}\rangle_{0}a_{kz}
+\alpha B_{iz}(1-\langle\sigma_{iz}^{2}\rangle_{0}) \, ,
\nonumber \\
(1+\alpha^{2})\langle \sigma_{iy}\sigma_{mx}\sigma_{kx}\sigma_{ly}\rangle_{0}\frac{\rmd}{\rmd t}2b_{kx,ly}=\langle B_{my}^{eff,0}\sigma_{iy}\sigma_{mz}\sigma_{kz}\rangle_{0}a_{kz}
\label{eqb} \\
\quad- \langle B_{mz}^{eff,0}\sigma_{iy}\sigma_{my}\sigma_{kz}\rangle_{0}a_{kz} - \langle B_{ix}^{eff,0}\sigma_{mx}\sigma_{iz}\sigma_{kz}\rangle_{0}a_{kz}
\nonumber \\
\quad+ \langle B_{iz}^{eff,0}\sigma_{mx}\sigma_{iz}\sigma_{kz}\rangle_{0}a_{kz}+ \alpha \langle B_{iy}^{eff,0}\sigma_{mx}\sigma_{kx}\sigma_{ly}\rangle_{0}2b_{kx,ly}
\nonumber \\
\quad+ \alpha \langle B_{mx}^{eff,0}\sigma_{iy}\sigma_{kx}\sigma_{ly}\rangle_{0}2b_{kx,ly} -\alpha \langle B_{i\alpha}^{eff,0}\sigma_{i\alpha}\sigma_{iy}\sigma_{mx}\sigma_{kx}\sigma_{ly}\rangle_{0}2b_{kx,ly}  \nonumber \\
\quad  -\alpha \langle B_{m\alpha}^{eff,0}\sigma_{m\alpha}\sigma_{iy}\sigma_{mx}\sigma_{kx}\sigma_{ly}\rangle_{0}2b_{kx,ly}-2\Sigma^{2}\langle \sigma_{iy}\sigma_{mx}\sigma_{kx}\sigma_{ly}\rangle_{0}2b_{kx,ly}
\nonumber \\
\quad-\delta_{im}\Sigma^{2}\langle \sigma_{iy}\sigma_{ix}\sigma_{kx}\sigma_{ly}\rangle_{0}2b_{kx,ly}-B_{mz}\langle\sigma_{iy}\sigma_{my}\rangle_{0}+B_{iz}\langle\sigma_{ix}\sigma_{mx}\rangle_{0} \, ,
\nonumber
\end{eqnarray}
where $b_{kx,ly}=b_{ly,kx}$ was assumed without the loss of generality, since the
antisymmetric part of this matrix doesn't contribute to the right-hand side of equation (\ref{Ppert}).

After solving the system of equations, the response of the system can be calculated as
\begin{eqnarray}
\langle \sigma_{iz} \rangle (t)=\langle\sigma_{iz}\sigma_{kz}\rangle_{0}a_{kz} (t) \, , \label{siz-exp}
\end{eqnarray}
which, as $a_{kz}(t)$, is linear in $B_{iz}$. In case of a periodic, finite wave vector external excitation, the lattice Fourier transform of the above quantity must be considered.

Next a mean field approach is introduced, where the unperturbed Hamiltonian $H^{0}(\{\boldsymbol{\sigma}_{i}\})$ in (\ref{H0}) is replaced by
\begin{eqnarray}
H_{0}^{\rm mf}(\{ \boldsymbol{\sigma}_i\})=4J_{xx}m \sum_i \sigma_{ix}+K_{x} \sum_i  \sigma_{ix}^{2} \, ,
\end{eqnarray}
where $m=\langle\sigma_{x}\rangle_{0,mf}$ has to be determined self-consistently. One has to assume that there is a finite but small $B_{x}$ external magnetic field, which chooses one of the degenerate states ($m>0$ or $m<0$) at low temperatures due to spontaneous symmetry breaking. We will assume the $m > 0$ case, but omit the $B_{x}\rightarrow 0+$ field in further calculations. Note that since $\langle\sigma_{y}\rangle_0^{\rm mf}=\langle\sigma_{z}\rangle_0^{\rm mf}=0$ holds due to the cylindrical symmetry of the system, the couplings $J_{yy}=J_{zz}$ vanish from the mean field Hamiltonian. Monte Carlo simulations indicate that the anisotropy together with the exchange leads to ferromagnetic ordering below a critical temperature. With the model parameters $J_{xx}=J_{yy}=J_{zz}=-1$ and $K_{x}=-0.1$, the critical temperature is $T_{c} \approx 0.7$, and this is only slightly changed by introducing Dzyaloshinsky-Moriya interactions of strength $D=0.1$. Since the mean field approximation underestimates the correlations of the system, it may be a suitable description only well below this critical temperature.

Selecting a single Fourier component in space and time for the perturbing field, $B_{jz}=\rme^{\rmi \omega t}\rme^{\rmi \boldsymbol{k}_{i}\cdot\boldsymbol{R_{j}}}B_{z}$, within the above mean field approximation, equations (\ref{eqa}) and (\ref{eqb}) become
\begin{eqnarray}
(1+\alpha^{2})\rmi \omega C_{zz}a=\big(B_{x}^{eff}-B_{y}^{eff}(\boldsymbol{k}_{i})\big)C_{xxyy}2b-\alpha B_{x}^{eff}C_{xxzz}a
\\
\qquad+\alpha B_{y}^{eff}(\boldsymbol{k}_{i})\big(C_{zz}-C_{yyzz}-C_{zzzz}\big)a-\Sigma^{2}C_{zz}a
+\alpha B_{z}(1-C_{zz}) \, ,
\nonumber \\
(1+\alpha^{2})\rmi \omega C_{xxyy}2b=-\big(B_{x}^{eff}-B_{y}^{eff}(\boldsymbol{k}_{i})\big)C_{xxzz}a
\\
\qquad +\alpha B_{x}^{eff}\big(C_{xxyy}-2C_{xxxxyy}\big)2b
\nonumber \\
\qquad +\alpha B_{y}^{eff}(\boldsymbol{k}_{i})\big(C_{xxyy}-2C_{xxyyyy}-2C_{xxyyzz}\big)2b
\nonumber \\
\qquad-3\Sigma^{2}C_{xxyy}2b+B_{z}(C_{xx}-C_{yy}) \, ,  \nonumber
\end{eqnarray}
where, for brevity,  the space and time Fourier components $a(\boldsymbol{k}_{i},\omega)$ and $b(\boldsymbol{k}_{i},\omega)$ are denoted by $a$ and $b$, respectively.  Furthermore, the notations $B_{x}^{eff}=-2K_{x}-4J_{xx}$, $B_{y}^{eff}(\boldsymbol{k}_{i})=B_{z}^{eff}(\boldsymbol{k}_{i})=(-J_{yy})(2\cos(\boldsymbol{k}_{i}\cdot\boldsymbol{\delta}_{1})+2\cos(\boldsymbol{k}_{i}\cdot\boldsymbol{\delta}_{2}))$ and $C_{\alpha\beta\dots}=\langle \sigma_{\alpha}\sigma_{\beta}\dots\rangle_0^{\rm mf}$ were used.

Introducing the quantities
\begin{eqnarray}
\omega_{0}(\boldsymbol{k}_{i})=B_{x}^{eff}-B_{y}^{eff}(\boldsymbol{k}_{i}),
\\
\lambda_{1}(\boldsymbol{k}_{i})=\alpha\Bigg[B_{x}^{eff}C_{xxzz}-B_{y}^{eff}(\boldsymbol{k}_{i})\big(C_{zz}-C_{yyzz}-C_{zzzz}\big)\Bigg]
\nonumber \\ +\Sigma^{2}C_{zz} \, ,
\\
\lambda_{2}(\boldsymbol{k}_{i})=\alpha\Bigg[B_{x}^{eff}\big(2C_{xxxxyy}-C_{xxyy}\big)-B_{y}^{eff}(\boldsymbol{k}_{i})\big(C_{xxyy}-2C_{xxyyyy} \nonumber
\\
\qquad-2C_{xxyyzz}\big)\Bigg]+3\Sigma^{2}C_{xxyy} \, ,
\end{eqnarray}
the lattice Fourier transform of the spin expectation value (\ref{siz-exp}) can be expressed as
\begin{eqnarray}
\langle\tilde{\sigma}_{z}\rangle(\boldsymbol{k}_{i},t)=C_{zz}\rme^{\rmi \omega t}a=\Bigg[\alpha(1-C_{zz})\rmi \omega+\frac{\alpha}{1+\alpha^{2}}\frac{1-C_{zz}}{C_{xxyy}}\lambda_{2}(\boldsymbol{k}_{i})
\label{szsol}  \\
\quad+\frac{1}{1+\alpha^{2}}(C_{xx}-C_{yy})\omega_{0}(\boldsymbol{k}_{i})\Bigg]\Bigg[-\omega^{2}+\frac{1}{(1+\alpha^{2})^{2}}\bigg(\omega_{0}^{2}(\boldsymbol{k}_{i})\frac{C_{xxzz}}{C_{zz}}
\nonumber \\
\quad+\frac{\lambda_{1}(\boldsymbol{k}_{i})\lambda_{2}(\boldsymbol{k}_{i})}{C_{xxyy}C_{zz}}\bigg)+\rmi 2\omega\frac{1}{1+\alpha^{2}}\frac{\lambda_{1}(\boldsymbol{k}_{i})C_{xxyy}+\lambda_{2}(\boldsymbol{k}_{i})C_{zz}}{2C_{xxyy}C_{zz}}\Bigg]^{-1}B_{z}\rme^{\rmi \omega t} \, .
\nonumber
\end{eqnarray}

Since the external excitation is a real-valued function, the calculations must be repeated for $B_{iz}=\rme^{-\rmi \omega t}\rme^{\rmi \boldsymbol{k}_{i}\cdot\boldsymbol{R_{i}}}B_{z}$, which will simply lead to the conjugate of expression (\ref{szsol}). One can also replace $\boldsymbol{k}_{i}$ with $-\boldsymbol{k}_{i}$, but this does not change the form of the expression, corresponding to the fact that the magnon spectrum is symmetric in the absence of Dzyaloshinsky-Moriya interaction. Taking the time average of the square of the sum of the $\omega$ and $-\omega$ components yields $\langle \big(2{\rm Re}\langle\tilde{\sigma}_{z}\rangle\big)^{2}(\boldsymbol{k}_{i})\rangle$, which is proportional to the absolute value squared of the right-hand side of (\ref{szsol}), in fact, a doubled Lorentzian function, just as in equation (\ref{curve}) for zero temperature. Calculating this expectation value makes it possible to determine the magnon energy (peak location) and the linewidth at finite temperatures.
Our numerical results will be shown and compared to simulations in \sref{simulations}. The basic effect is that the magnon energy decreases with increasing temperature, while the linewidth increases.

\subsection{Finite temperature effects: the variational method\label{sec:vm}}

Another way of determining the magnon energies at finite temperatures is a variational method based on a quantummechanical treatment first described by Bloch\cite{Bloch} for an isotropic Heisenberg model on a cubic lattice. The method was extended to include on-site and two-site anisotropies as well as different lattice types\cite{Rastelli}. In this paper the method is extended to also include Dzyaloshinsky-Moriya interactions. For the Hamiltonian again
nearest-neighbour exchange interactions with $J_{yy}=J_{zz}$  and next-nearest-neighbour Dzyaloshinsky-Moriya interactions were assumed,
\begin{eqnarray}
H(\{\boldsymbol{\sigma}_{i}\})=J_{yy}\sum_{\langle i,j \rangle_{1}} \boldsymbol{\sigma}_{i}\boldsymbol{\sigma}_{j}+ \big(J_{xx}-J_{yy}\big)\sum_{\langle i,j \rangle_{1}} \sigma_{ix}\sigma_{jx}+ K_{x}\sum_{i}\sigma_{ix}^{2} \nonumber
\\
\qquad+\sum_{\langle i,j \rangle_{2}} D^{ij}\Big(\sigma_{iy}\sigma_{jz}-\sigma_{iz}\sigma_{jy}\Big) \, ,
\end{eqnarray}
Treating $\sigma_{i\alpha}$ as spin operators, a bosonic representation can be introduced in terms of the Dyson-Maleev transformation\cite{Dyson,Maleev},
\begin{eqnarray}
\sigma_{ix}=S-a_{i}^{\dag}a_{i} \, ,
\\
\sigma_{i}^{+}=\sigma_{iy}+\rmi \sigma_{iz}=\sqrt{2S}\Big(1-\frac{a_{i}^{\dag}a_{i}}{2S}\Big)a_{i} \, ,
\\
\sigma_{i}^{-}=\sigma_{iy}-\rmi \sigma_{iz}=\sqrt{2S}a_{i}^{\dag} \, ,
\end{eqnarray}
where the $x$ axis was used as the primary quantization axis, therefore the bosonic vacuum corresponds to the ground state of the spin system. It should be emphasized that only the physical part of the bosonic system, $0 \leq a_{i}^{\dag}a_{i} \leq 2S$, has to be considered.

The Hamiltonian written in the bosonic operators describes an interacting system, as it contains terms including the products of two and four bosonic operators. The variational approach is based on calculating the free energy,
\begin{eqnarray}
F=\langle H \rangle -TS \, ,
\end{eqnarray}
where the expectation value is a thermal average taken with respect to the eigenstates of a suitable non-interacting model Hamiltonian
\begin{eqnarray}
H^{0}=\sum_{\boldsymbol{k}}\omega_{\boldsymbol{k}}(T)a_{\boldsymbol{k}}^{\dag}a_{\boldsymbol{k}} \, ,
\end{eqnarray}
with $a_{\boldsymbol{k}}=\frac{1}{\sqrt{N}}\sum_{i}\rme^{\rmi \boldsymbol{k}\cdot\boldsymbol{R}_{i}}a_{i}$, the Fourier transform of the bosonic operators, and $\omega_{\boldsymbol{k}}(T)$ is a temperature-dependent quasiparticle energy, to be determined later by minimization of the free energy. In the classical limit, the expectation value of the original Hamiltonian takes the form,
\begin{eqnarray}
\langle H \rangle=-4J_{yy}\sum_{\boldsymbol{k}}\big(1-\gamma_{\boldsymbol{k}}^{(1)}\big)n_{\boldsymbol{k}}(T)-4\big(J_{xx}-J_{yy}\big)
\sum_{\boldsymbol{k}}n_{\boldsymbol{k}}(T) \nonumber
\\
\qquad-2K_{x}\sum_{\boldsymbol{k}}n_{\boldsymbol{k}}(T)-2D\sum_{\boldsymbol{k}}\gamma_{\boldsymbol{k}}^{(2)}n_{\boldsymbol{k}}(T) \nonumber
\\
\qquad+2J_{yy}\frac{1}{N}\sum_{\boldsymbol{k},\boldsymbol{k}'}\big(1+\gamma_{\boldsymbol{k}-\boldsymbol{k}'}^{(1)}
-2\gamma_{\boldsymbol{k}'}^{(1)}\big)n_{\boldsymbol{k}}(T)n_{\boldsymbol{k}'}(T)\nonumber
\\
\qquad+2\big(J_{xx}-J_{yy}\big)\frac{1}{N}\sum_{\boldsymbol{k},\boldsymbol{k}'}\big(1+\gamma_{\boldsymbol{k}-\boldsymbol{k}'}^{(1)}\big)
n_{\boldsymbol{k}}(T)n_{\boldsymbol{k}'}(T)\nonumber
\\
\qquad+2K_{x}\frac{1}{N}\sum_{\boldsymbol{k},\boldsymbol{k}'}n_{\boldsymbol{k}}(T)n_{\boldsymbol{k}'}(T)+2D\frac{1}{N}
\sum_{\boldsymbol{k},\boldsymbol{k}'}\gamma_{\boldsymbol{k}'}^{(2)}n_{\boldsymbol{k}}(T)n_{\boldsymbol{k}'}(T) \, ,
\end{eqnarray}
where $D$ is the magnitude of the Dzyaloshinsky-Moriya vector, $\gamma_{\boldsymbol{k}}^{(1)}=\cos(\frac{\sqrt{2}}{2}k_{x}a)\cos(\frac{1}{2}k_{y}a)$ and $\gamma_{\boldsymbol{k}}^{(2)}=\sin(k_{y}a)$ are geometrical factors characteristic for the lattice, and $n_{\boldsymbol{k}}(T)=\langle a_{\boldsymbol{k}}^{\dag}a_{\boldsymbol{k}} \rangle$ is the occupation number. Also in the classical limit the Boltzmann entropy
\begin{eqnarray}
S=\sum_{\boldsymbol{k}}\ln\Big(n_{\boldsymbol{k}}(T)\Big)
\end{eqnarray}
is considered instead of the entropy of a non-interacting bosonic system. The quasiparticle energies and the occupation numbers are therefore related to each other through
\begin{equation}
\omega_{\boldsymbol{k}}(T)=\frac{T}{n_{\boldsymbol{k}}(T)} \, .\label{scons1}
\end{equation}
Requiring that the occupation numbers $n_{\boldsymbol{k}}(T)$ minimize the free energy $F$, leads to the set of equations
\begin{eqnarray}
\omega_{\boldsymbol{k}}(T)=-4J_{yy}\big(1-\gamma_{\boldsymbol{k}}^{(1)}\big)-4\big(J_{xx}-J_{yy}\big)-2K_{x}-2D\gamma_{\boldsymbol{k}}^{(2)}\nonumber
\\
\qquad+4J_{yy}\frac{1}{N}\sum_{\boldsymbol{k}'}\big(1+\gamma_{\boldsymbol{k}-\boldsymbol{k}'}^{(1)}-\gamma_{\boldsymbol{k}'}^{(1)}-
\gamma_{\boldsymbol{k}}^{(1)}\big)n_{\boldsymbol{k}'}(T)\nonumber
\\
\qquad+4\big(J_{xx}-J_{yy}\big)\frac{1}{N}\sum_{\boldsymbol{k}'}\big(1+\gamma_{\boldsymbol{k}-\boldsymbol{k}'}^{(1)}\big)n_{\boldsymbol{k}'}(T)\nonumber
\\
\qquad+4K_{x}\frac{1}{N}\sum_{\boldsymbol{k}'}n_{\boldsymbol{k}'}(T)+2D\frac{1}{N}\sum_{\boldsymbol{k}'}\big(\gamma_{\boldsymbol{k}'}^{(2)}+
\gamma_{\boldsymbol{k}}^{(2)}\big)n_{\boldsymbol{k}'}(T) \, .\label{scons2}
\end{eqnarray}

Equations (\ref{scons1}) and (\ref{scons2}) can be used to self-consistently determine the occupation numbers and the temperature-dependent magnon energies. It is easy to see that at $T=0$, (\ref{scons2}) simplifies to the magnon spectrum in (\ref{spekeq}). The magnon energies decrease with increasing temperature since all the corrections from the interacting part of the Hamiltonian have a different sign compared to the noninteracting part. This method obviously does not take into account the phenomenological Gilbert damping, $\alpha$, which slightly modifies the magnon energies in (\ref{curve}) and (\ref{szsol}), but this effect is small if $\alpha \ll 1$, which is usual for a ferromagnetic system. On the other hand, this approximation does not give any information on the linewidth of the response function. Although it is possible to determine this quantity using different quantumtheoretical descriptions\cite{Lutsev}, we do not discuss such an approach here, since in the quasiclassical limit the Gilbert damping is responsible for the linewidth, similar to as given in (\ref{curve}) at $T=0$ K. The numerical results from equations (\ref{scons1}) and (\ref{scons2}) will also be compared to simulations later on.

\section{Atomistic spin dynamics simulations}

\subsection{Simulations at zero temperature}

To confirm the theoretical results discussed in the above sections, atomistic spin dynamics simulations were carried out. The code we developed solves the stochastic Landau-Lifshitz-Gilbert equations (\ref{LLG}) on a two-dimensional lattice, using the stochastic Heun method with the symmetry-preserving modifications described in \cite{Mentink}. According to the previous calculations, in the simulations the geometry of a $(110)$ surface of a bcc lattice was considered and a ferromagnetic Heisenberg model with nearest-neighbour coupling of unit strength: $J=-1$. Choosing the energy scale also determines a timescale, but during a simulation where only the stationary properties of the system are examined, this timescale is only important to determine the length of the transients that should be omitted from the calculation of averages. An anisotropy constant $K_{x}<0$ was also used that reinforced a ground state ferromagnetic order with all spins parallel to the $x$ axis. This basic model was then extended by other coupling coefficients, namely, next nearest neighbor Dzyaloshinsky-Moriya interactions ($D$) or different diagonal elements in the $\boldsymbol{J}_{ij}$ tensor ($J_{xx}, J_{yy}, J_{zz}$) describing two-site anisotropy. The simulations were performed on a lattice of $32\times32$ atoms with periodic boundary conditions. As starting configuration the ferromagnetic ground state was chosen.

A time-dependent, inhomogeneous field $2B_{z}\cos\big(\boldsymbol{k}_{i}\cdot\boldsymbol{r}\big)\cos\big(\omega t\big)$ was chosen as the external excitation, where $\boldsymbol{k}_{i}$ is a wave vector in the first Brillouin zone. The response of the system was calculated as $S(\boldsymbol{k}_{i},\omega)=\langle m_{y}^{2}(\boldsymbol{k}_{i}) \rangle + \langle m_{z}^{2}(\boldsymbol{k}_{i}) \rangle - \langle m_{y}(\boldsymbol{k}_{i}) \rangle^{2} - \langle m_{z}(\boldsymbol{k}_{i}) \rangle^{2}$, where $\langle \boldsymbol{m}(\boldsymbol{k}_{i}) \rangle=\langle \sum_{j} \cos{(\boldsymbol{k}_{i}\cdot\boldsymbol{R}_{j})} \boldsymbol{\sigma}_{j}\rangle$, and $\langle \rangle$ stands for time averaging. The difference to the expression (\ref{S}) is that the simulation code uses the Descartes components of the spins instead of the angle variables in the linearized equations (\ref{eq1}) and (\ref{eq2}), but close to the ground state these variables are basically identical ($\sigma_{iy} \approx \beta_{2i}, \sigma_{iz} \approx -\beta_{1i}$). In order to obtain a resonance curve, $S(\boldsymbol{k}_{i},\omega)$ was calculated for different values of the $\omega$ frequency. The value of the $B_{z}$ amplitude had to be chosen carefully, since at large values it may move the spins very far from the ferromagnetic ground state, and the system may become disordered, however, for small values of $B_{z}$ the resonance curve is hardly noticeable over the thermal background.

Choosing the correct value for the Gilbert damping, $\alpha$, was also important. On the one hand, large damping increases the half-width of the resonance curve, meaning that the frequency can be changed in larger steps, and more importantly, making the transients decay faster so shorter simulation times will suffice. On the other hand, the magnon softening due to the Dzyaloshinsky-Moriya interaction and the different diagonal coupling coefficients is almost independent of the damping. This shift in the magnon energies can be easier detected in the case of sharper peaks that can be obtained with smaller $\alpha$.

In case of lattice defects, the initial configuration of the simulation was changed by creating small connected droplets of zero spins in the lattice. The method of the simulation was the same as in the case of the perfect lattice, namely, that a time dependent $B_{z}$ field of $\omega$ angular frequency was used, and the response of the system, $S(\boldsymbol{0},\omega)=\langle m_{y}^{2} \rangle + \langle m_{z}^{2} \rangle - \langle m_{y} \rangle^{2} - \langle m_{z} \rangle^{2}$, was calculated as a function of $\omega$. In this case ten droplets of zero spins were chosen in the $32\times32$ size lattice, each containing $5$ sites, giving a total number of $50$ vacancies. Of course, here only the nonzero spins are used in calculating the averages, the ones set to zero are omitted.

\begin{figure}[H]
\centering
\includegraphics[width=12cm,height=7.8cm]{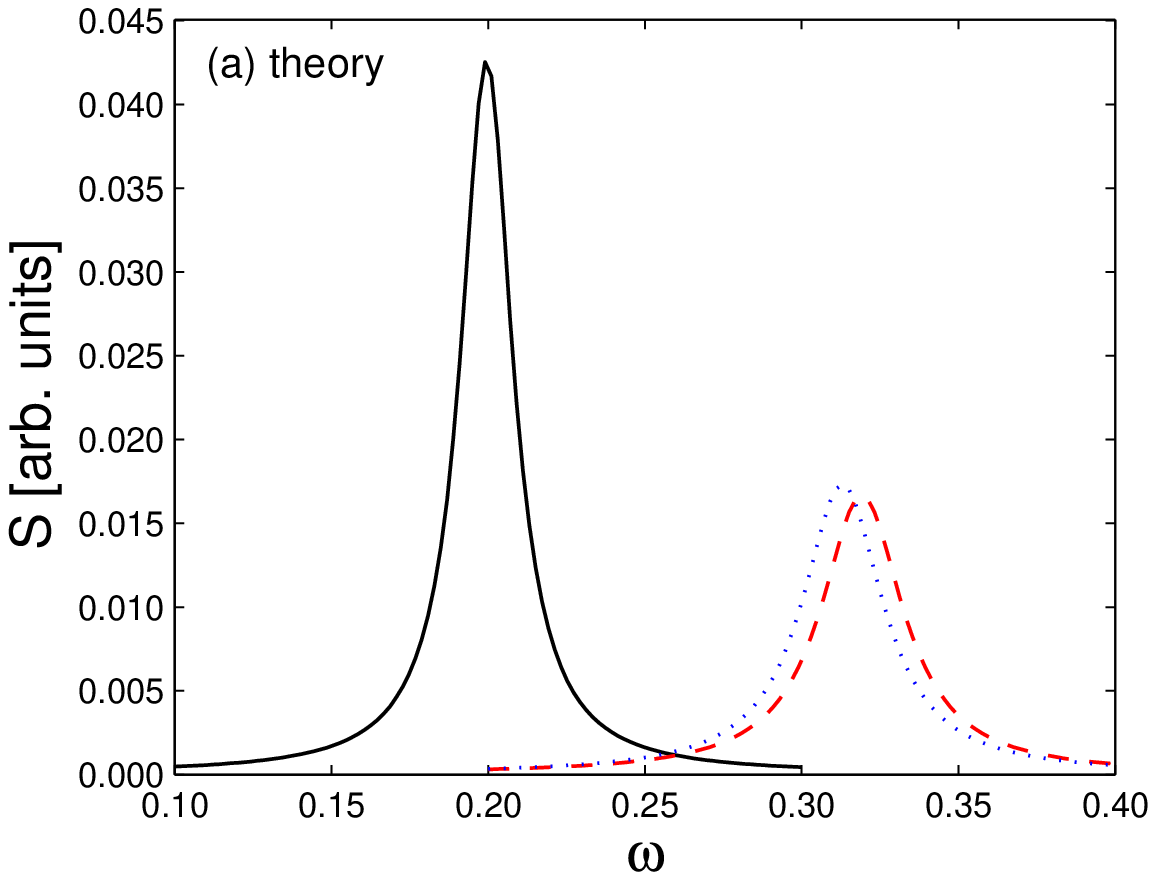}
\includegraphics[width=12cm,height=7.8cm]{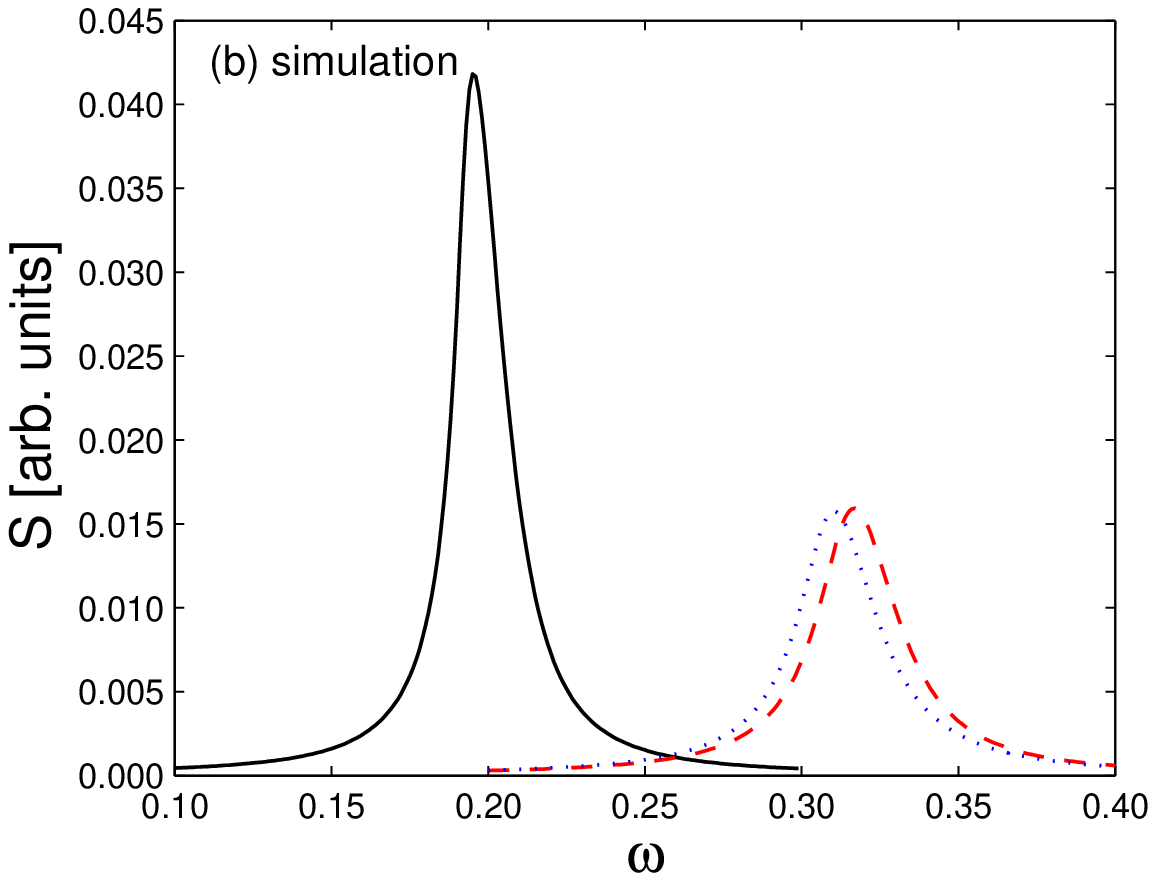}
\caption{Response functions for $\boldsymbol{k}_{i}=\boldsymbol{0}$  (a) based on the theory presented in section \ref{lattdef} and (b) obtained from numerical simulations for the system with isotropic ($J_{xx}=J_{yy}=J_{zz}=-1$, solid line) and anisotropic ($J_{xx}=-1.02, J_{yy}=J_{zz}=-0.99$, dashed line) exchange interactions. The other model parameters are $\alpha=0.05$, $T=0$, $K_{x}=-0.1$, $n=1024$. The result of the lattice defect model is also shown for the anisotropic Hamiltonian, where the value of the spin vectors is set to zero at $m=50$ different lattice points (dotted line). The overall features of the peaks on the two panels are in very good agreement.}
\label{D0D0J}
\end{figure}

Figure \ref{D0D0J} shows the results of the linear approximation as well as the simulations at $\boldsymbol{k}_{i}=\boldsymbol{0}$. Without Dzyaloshinsky-Moriya interactions, the response of the system can be modelled by a single resonance curve. The peak is located approximately at $2|K_{x}|=0.2$, i.e. at the energy of the zero wave vector magnon as indicated by equation (\ref{spekeq}). Introducing different diagonal coupling coefficients increases this value by about $-4(J_{xx}-J_{yy})=0.12$. The simulations were carried out also for finite Dzyaloshinsky-Moriya interactions ($D=0.1$), but this caused no detectable difference in the obtained resonance curves, in agreement with the linear approximation (\ref{curve}). By adding lattice defects to the system, the magnon energy decreases if two-site anisotropy is present. If the Hamiltonian only contains on-site anisotropy, the results of the simulation show no difference compared to the case of the perfect lattice.


\begin{figure}[H]
\centering
\includegraphics[width=12cm,height=7.8cm]{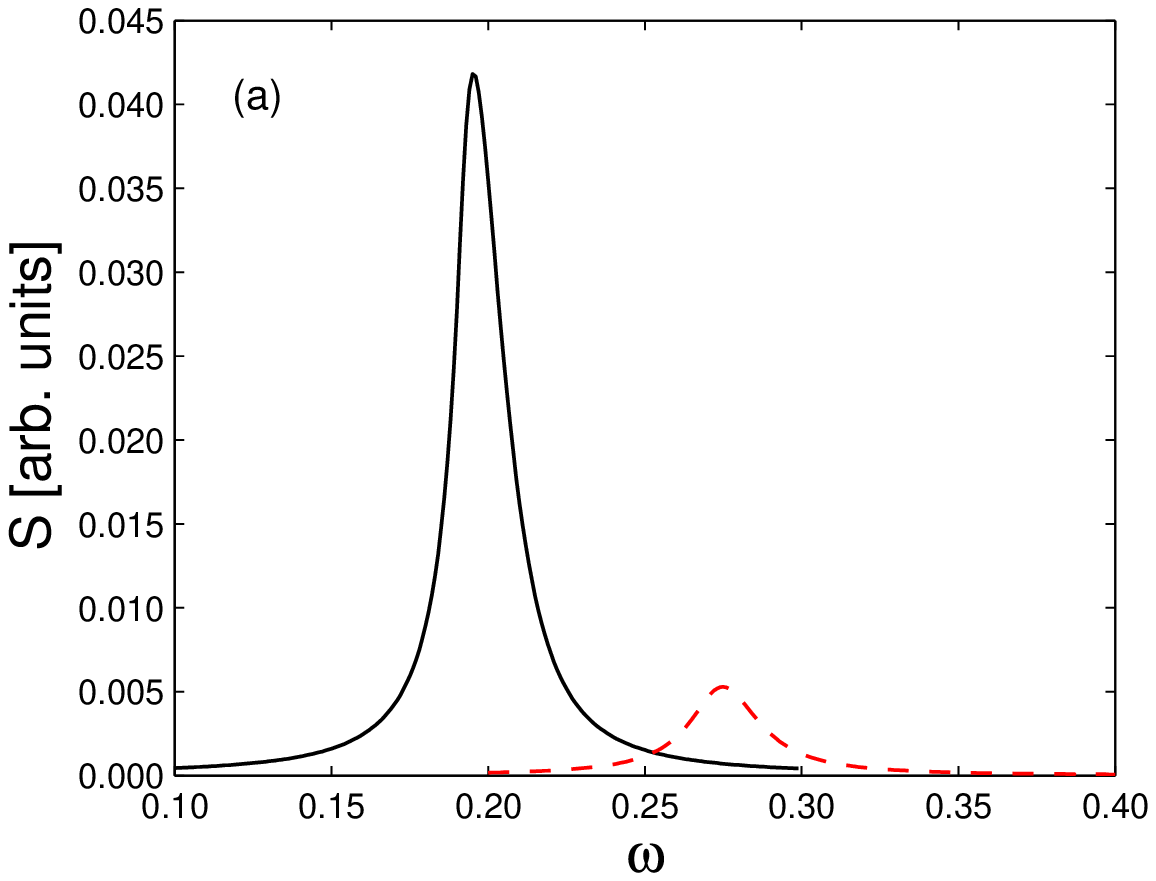}
\includegraphics[width=12cm,height=7.8cm]{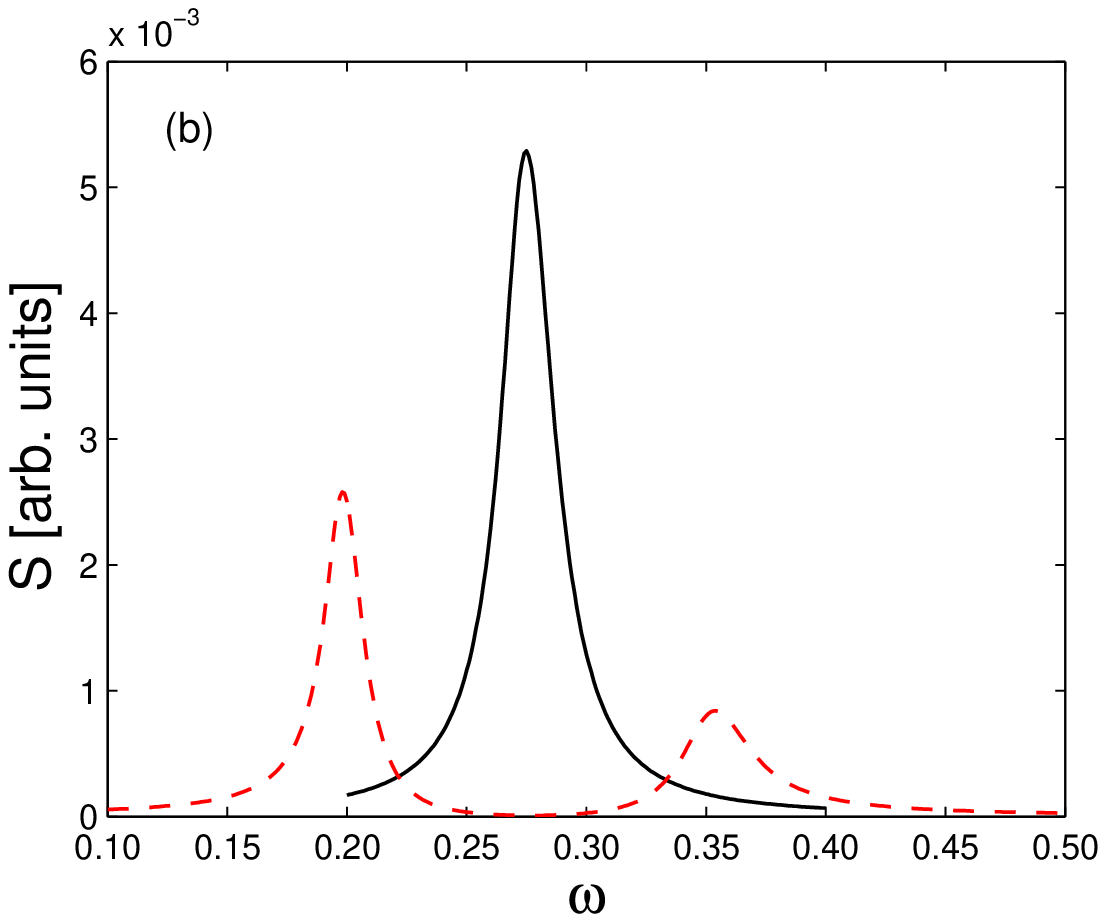}
\caption{Simulated response functions at $T=0$. Panel (a) shows the response without Dzyaloshinsky-Moriya interaction at  $\boldsymbol{k}_{i}=\boldsymbol{0}$ (solid line) and at $(k_{x},k_{y})=(0,\frac{2\pi}{16a})$ (dashed line). Panel (b) displays the response at $(k_{x},k_{y})=(0,\frac{2\pi}{16a})$ without (solid line) and with Dzyaloshinsky-Moriya interaction, $D=0.1$ (dashed line). The other parameters are $J_{xx}=J_{yy}=J_{zz}=-1$, $\alpha=0.05$ and $K_{x}=-0.1$. The locations and splitting of the peaks are in good agreement with the linear approximation (\ref{spekeq}).}
\label{D0D01ky0ky01}
\end{figure}

The results of the simulations for finite wave vector excitations can be seen in figure \ref{D0D01ky0ky01}. Here the wave vector points in the $y$ direction, since (\ref{spekeq}) suggests that, in the presence of Dzyaloshinsky-Moriya interactions, the degeneracy between the $\boldsymbol{k}_{i}$ and $-\boldsymbol{k}_{i}$ magnon energies is lifted for wave vectors along this direction. The value for $k_{y}$ is $\sqrt{2}k_y=2\pi\frac{\sqrt{2}}{16a}$, $a$ being the lattice constant, since in a $32\times32$ lattice this is the smallest allowed wave vector. Figure \ref{D0D01ky0ky01}(a) demonstrates that without Dzyaloshinsky-Moriya interactions there remains a single peak, however, it is shifted to a higher energy for finite wave vector. The value of the shift is consistent with the linear approximation, $4|J_{yy}|(1-\cos{\frac{1}{2}ak_{y}})=4|J_{yy}|(1-\cos{\frac{2\pi}{32}})\approx 0.077$. In the presence of the Dzyaloshinsky-Moriya interaction, figure \ref{D0D01ky0ky01}(b) shows the expected splitting of the peak, in agreement with (\ref{curve}). Even the value of the splitting is in good agreement with the result of the linear approximation, $4D\sin{\frac{2\pi}{16}}\approx 0.153$.

\begin{figure}[H]
\centering
\includegraphics[width=12cm,height=7.5cm]{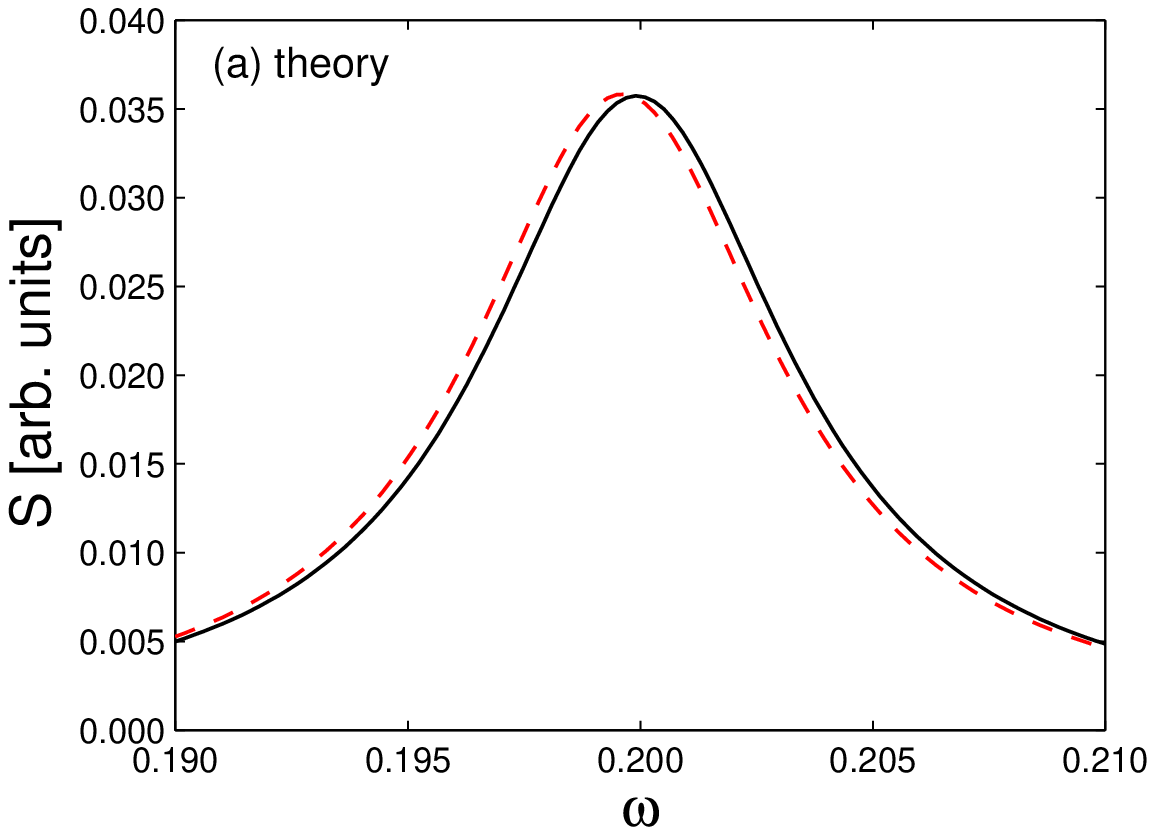}
\includegraphics[width=12cm,height=7.5cm]{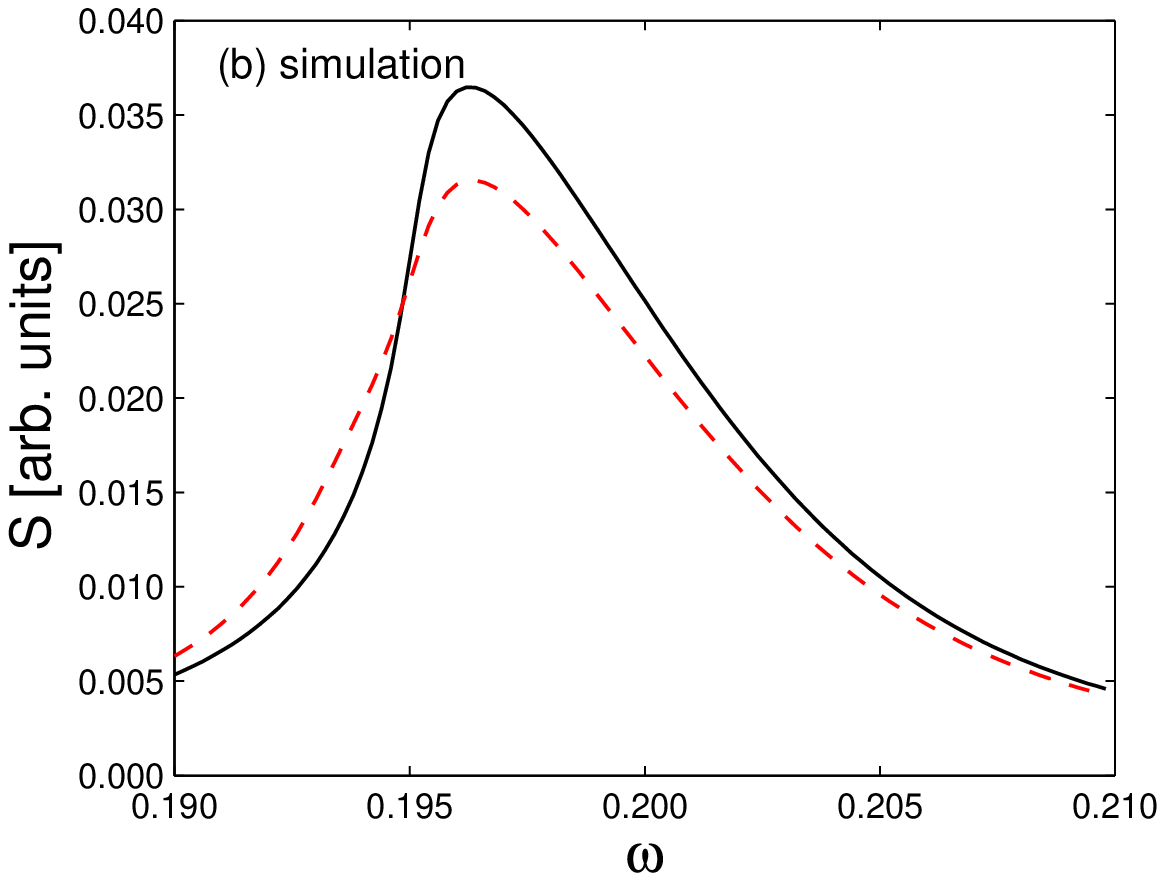}
\caption{Response functions (a) based on the theory in section \ref{lattdef} and (b) from numerical simulations  at $T=0$ and $\boldsymbol{k}_{i}=\boldsymbol{0}$ in the presence of Dzyaloshinsky-Moriya interaction, $D=0.1$, 
for a perfect lattice (solid line) and for an imperfect lattice (dashed line). The model parameters are $J_{xx}=J_{yy}=J_{zz}=-1$, $\alpha=0.02$, $K_x=-0.1$, $n=1024$ and $m=50$. Note the asymmetric lineshape of the simulated response functions.}
\label{D0D01lyuk50}
\end{figure}

Figure \ref{D0D01lyuk50}(b) justifies that in a disordered system, the Dzyaloshinsky-Moriya interaction does indeed change the shape of the curve, while it has no effect in a perfect lattice at $\boldsymbol{k}_{i}=\boldsymbol{0}$. Since this effect is expected to be smaller than the one due to the anisotropy of the diagonal coupling coefficients, here a smaller value of Gilbert damping constant, $\alpha=0.02$, is used,  instead of $\alpha=0.05$, making the peaks sharper. It is observed that both the chiral interaction and the lattice defects are necessary to change the lineshape, just as in the theoretical calculations in section \ref{lattdef} and as seen in figure \ref{D0D01lyuk50}(a). The peak is shifted towards lower magnon energies in the simulation as well as in the calculation. This effect is, however, somewhat ambiguous for the latter case, since decreasing the Gilbert damping in the simulation makes the lineshape more asymmetric, whereas this feature is absent in the theoretical calculations. Nevertheless, the peak becomes wider with vacancies in the lattice, and the maximum of the peak decreases. Importantly, this effect is practically unchanged when the spins are set to zero at different sets of lattice points.

In spin dynamics simulations it is useful to calculate the spin-spin correlation function\cite{Chen},
\begin{eqnarray}
C_{\alpha}\big(\boldsymbol{R}_{i}-\boldsymbol{R}_{j},t_{n}\big)=\langle \sigma_{i\alpha}(t_{n})\sigma_{j\alpha}(0) \rangle - \langle \sigma_{i\alpha}(t_{n}) \rangle \langle \sigma_{j\alpha}(0) \rangle \, ,
\end{eqnarray}
where $\alpha=x, y, z$ and $\langle \rangle$ stands for ensemble average. Here it is denoted explicitly that the correlation function is only calculated at discrete time points $t_{n}$, which will lead to a finite maximal frequency. This average is achieved by starting the simulation multiple times from the same initial configuration, but using different seeds of the random number generator, leading to different trajectories in the phase space.

The dynamic structure factor is defined as
\begin{eqnarray}
S_{\alpha}\big(\boldsymbol{k}_{i},\omega_{j}\big)=\sum_{l,m,n}\rme^{\rmi\boldsymbol{k}_{i}\cdot(\boldsymbol{R}_{l}-\boldsymbol{R}_{m})}e^{i\omega_{j} t_{n}} C_{\alpha}\big(\boldsymbol{R}_{l}-\boldsymbol{R}_{m},t_{n}\big) ,
\end{eqnarray}
which is the Fourier transform of the correlation function, discretized in space and time, but also containing a double summation over the lattice points, which corresponds to a lattice averaging besides the Fourier transformation. Due to the finite simulation time, we only get the value of the dynamic structure factor at discrete $\omega_{j}$ frequencies, while the finite size of the lattice leads to a discretization in the momentum space. Using the space Fourier transform $\hat{\sigma}_{\alpha}(\boldsymbol{k}_{i},t_{n})=\sum_{l}\rme^{\rmi\boldsymbol{k}_{i}\cdot\boldsymbol{R}_{l}}\sigma_{l\alpha}(t_n)$, as well as the space and time Fourier transform $\tilde{\sigma}_{\alpha}(\boldsymbol{k}_{i},\omega_{j})=\sum_{n}\rme^{\rmi\omega_{j} t_{n}}\hat{\sigma}_{\alpha}(\boldsymbol{k}_{i},t_{n})$, of the spin components, the dynamic structure factor can be rewritten as
\begin{eqnarray}
S_{\alpha}\big(\boldsymbol{k}_{i},\omega_{j}\big)=\langle \tilde{\sigma}_{\alpha}(\boldsymbol{k}_{i},\omega_{j})\hat{\sigma}_{\alpha}(\boldsymbol{k}_{i},0) \rangle - \langle \tilde{\sigma}_{\alpha}(\boldsymbol{k}_{i},\omega_{j}) \rangle \langle \hat{\sigma}_{\alpha}(\boldsymbol{k}_{i},0) \rangle \, .
\end{eqnarray}

Calculating the time Fourier transform of the spin components is basically the same as calculating the linear response of the system to an external excitation with time dependence $\rme^{\rmi \omega t}$, as long as the system is close to the ground state and the linear approximation is valid. From the complex $y$ and $z$ components of the dynamic structure factor, the linear response can be expressed at discrete frequencies as
\begin{eqnarray}
S\big(\boldsymbol{k}_{i},\omega_{j}\big)=|S_{y}\big(\boldsymbol{k}_{i},\omega_{j}\big)|^{2}+|S_{z}\big(\boldsymbol{k}_{i},\omega_{j}\big)|^{2},
\label{S3}
\end{eqnarray}
a suitable real quantity for describing the deviation of the system from its ferromagnetic ground state. For a given $\boldsymbol{k}_{i}$ wave vector, $S(\boldsymbol{k}_{i},\omega_{j})$ is expected to have a similar form to the theoretical results, namely, equation (\ref{curve}) at $T=0$ and equation (\ref{szsol}) at finite temperature.

\begin{figure}[H]
\centering
\includegraphics[width=12cm,height=7.5cm]{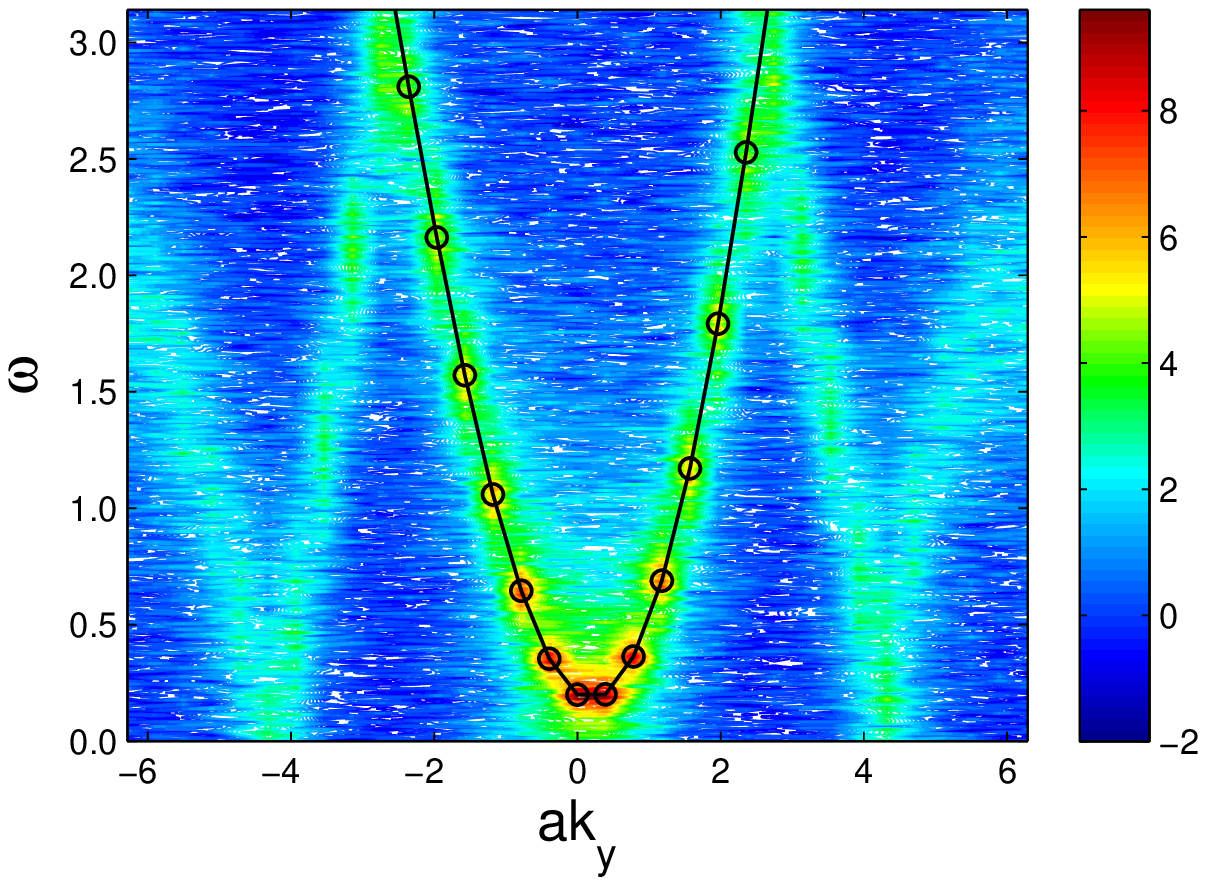}
\caption{Contour plot of calculated dynamic structure factor (logarithmic scale), $S(\boldsymbol{k}_{i},\omega_{j})$, see equation (\ref{S3}), as a function of the $k_{y}$ component of the wave vector as well as the angular frequency, with the parameters  $J_{xx}=J_{yy}=J_{zz}=-1$, $D=0.1$, $K_{x}=-0.1$, $\alpha=0.05$, $T=0.01$, $k_{x}=0$. The time delay for calculating the spin values was set to $\Delta t=1$, corresponding to a maximal frequency of $\omega_{max}=\pi$. The resolution in frequency is $\Delta\omega=\frac{2\pi}{1000}$, because the length of the examined time interval was $t_{max}=1000$. The values for the wave vector are $k_{y}=j\frac{2\pi}{16a}$, where $a$ is the lattice constant and $j$ is an integer between $-16$ and $16$ because of the lattice size $32\times32$. At high wave vectors, the magnon energies are higher than $\omega=\pi$, but these values appear mirrored due to the discretization in time. The open circles represent the magnon spectrum calculated from the linear approximation at $T=0$, equation (\ref{spekeq}).}
\label{D01contour}
\end{figure}

For the case of isotropic coupling ($J=-1$) and presence of on-site anisotropy ($K_{x}=-0.1$) as well as of Dzyaloshinsky-Moriya interaction ($D=0.1$), figure \ref{D01contour} shows $S(\boldsymbol{k}_{i},\omega_{j})$ as a function of the wave vector along the $y$ direction and the angular frequency.  The magnon excitations correspond to peaks in the $k_{y}-\omega$ plane. In these simulations no exciting external field was applied, since the excitations of the spin system can appear solely due to the finite temperature ($T=0.01$). Importantly, the theoretically calculated magnon dispersion based on equation (\ref{spekeq}) is in fairly good agreement with the peaks of the dynamic structure factor obtained from simulations.

\begin{figure}[H]
\centering
\includegraphics[width=12cm,height=7.5cm]{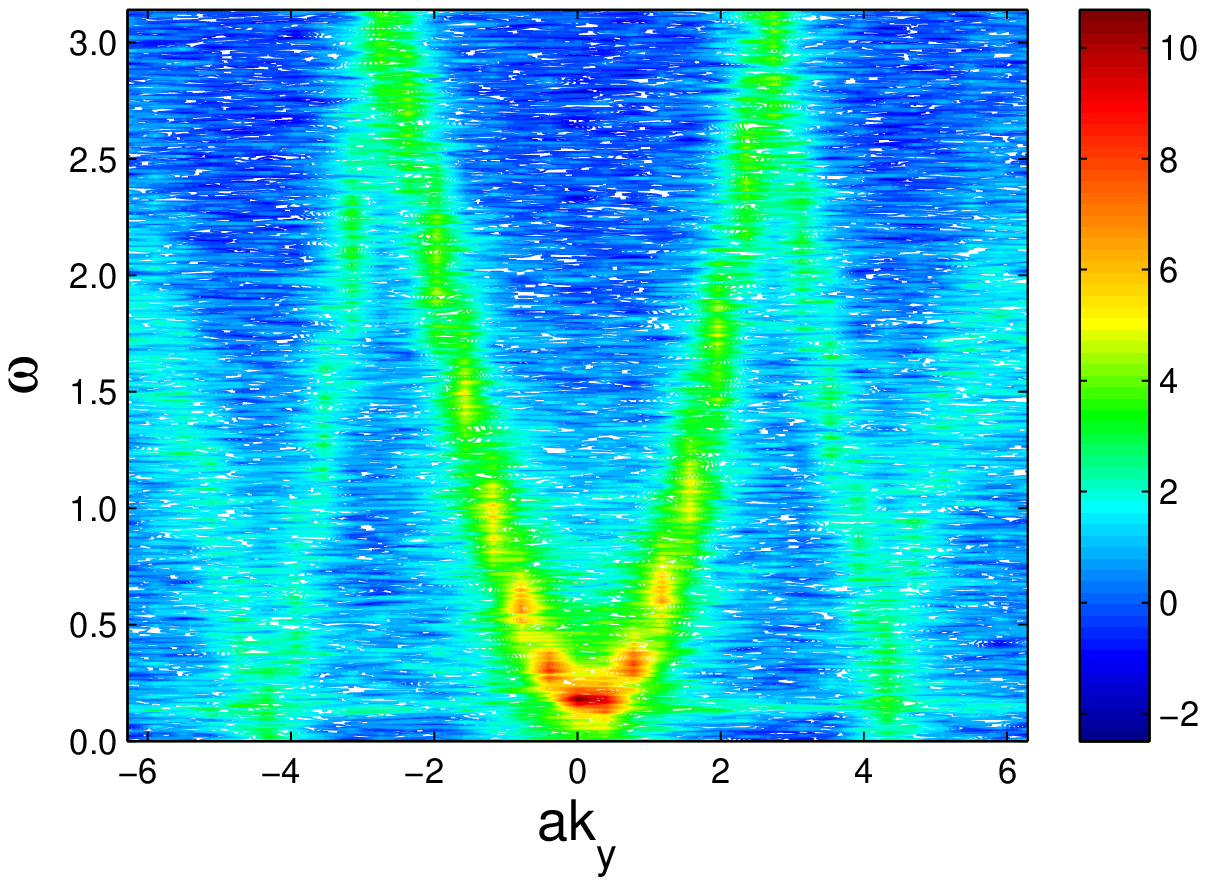}
\caption{Contour plot of calculated dynamic structure factor (logarithmic scale), $S(\boldsymbol{k}_{i},\omega_{j})$ as a function of the $k_{y}$ component of the wave vector as well as the angular frequency, when the value of the spins is set to zero at $50$ different lattice points. The parameters of the simulations were $J_{xx}=J_{yy}=J_{zz}=-1$, $D=0.1$, $K_{x}=-0.1$, $\alpha=0.05$, $T=0.01$, $k_{x}=0$, and an external excitation $B_{z}$ was applied, with zero wave vector and angular frequency of $0.2$. Note the horizontal stripe at $\omega=0.2$ corresponding to magnon scattering.}
\label{lyuk50contour}
\end{figure}

Figure \ref{lyuk50contour} shows the effect of lattice defects on the dynamic structure factor. Here a homogeneous external excitation $B_z$ was applied with an angular frequency $\omega=0.2$, since it makes it easier to realize the difference between the contour diagram with and without lattice defects. Since the quasimomentum is not conserved, a zero wave vector excitation will be scattered and will appear at other wave vectors at the same frequency. This is indicated as a horizontal stripe in the figure at $\omega=0.2$. As before, thermal excitations are also scattered to finite wave vectors, but the corresponding stripes (peak positions) are more difficult to notice since the original peaks are also smaller due to lattice defects.

\subsection{Simulations at finite temperature\label{simulations}}

We determined the linear response of the system, $S(\boldsymbol{k}_{i},\omega_{j})$, from simulations at finite temperatures as well. Lorentzian functions were fitted to the resonance curves, and the fitting parameters for the location and the half-width of the peaks were compared to the theoretical predictions from the mean field approach (\ref{szsol}) and the variational method (\ref{scons1})-(\ref{scons2}).

\begin{figure}[H]
\centering
\includegraphics[width=12cm,height=7.5cm]{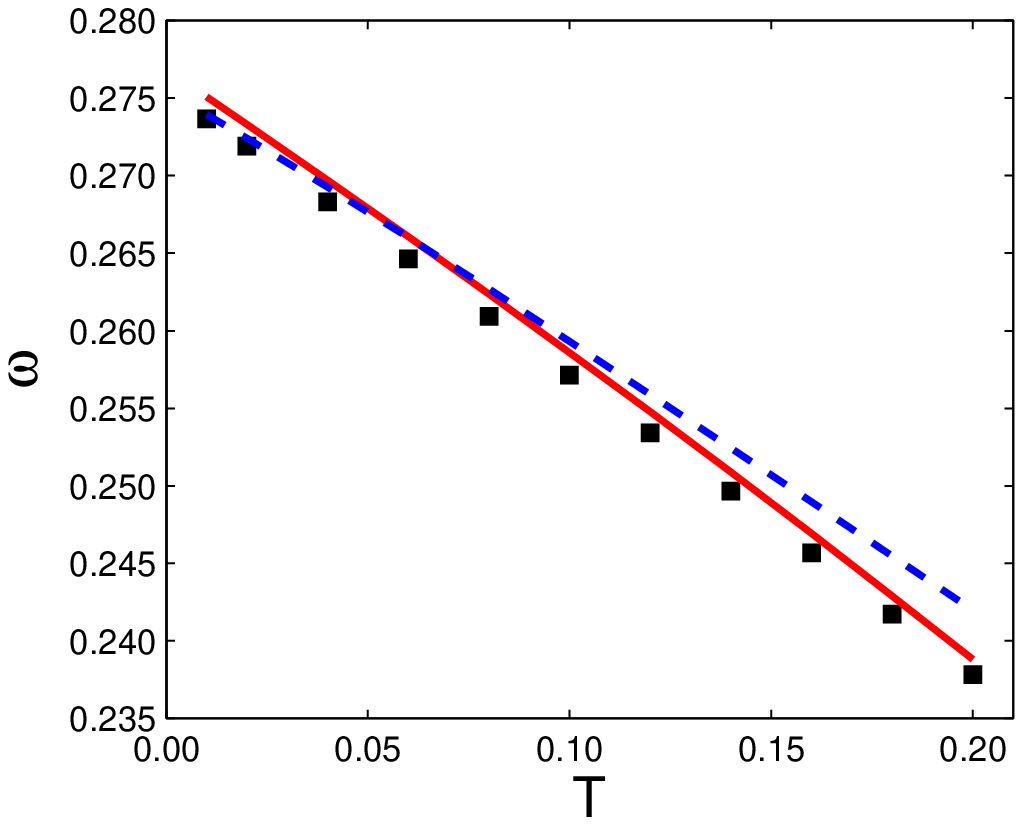}
\caption{Calculated magnon frequencies at $(k_{x},k_{y})=(0,\frac{2\pi}{16a})$ for temperatures $0.01 \le T \le 0.2$. The results of the simulations (squares) are compared to those from the the mean field approach (dashed line) and from the variational method (solid line), see sections \ref{sec:mf} and \ref{sec:vm}. The model parameters are $J_{xx}=J_{yy}=J_{zz}=-1$, $D=0$, $K_{x}=-0.1$ and $\alpha=0.05$.}
\label{D0ky01Txc}
\end{figure}

Figure \ref{D0ky01Txc} depicts the magnon frequency as a function of temperature at a given wave vector. Both the mean field and the variational methods are in good qualitative agreement with the simulations as they reproduce the decrease of the magnon frequency with increasing temperature. Due to the missing Gilbert damping, in the case of the variational method, a seemingly constant shift compared to the values from the simulations can be inferred from the figure. In the mean field approximation, the slope of the curve is slightly different from that of the simulations. The decreasing energy of the magnons may be explained by the decreasing magnetization, which is also a linear function of the temperature in the classical Heisenberg model at low temperatures.

\begin{figure}[H]
\centering
\includegraphics[width=12cm,height=7.5cm]{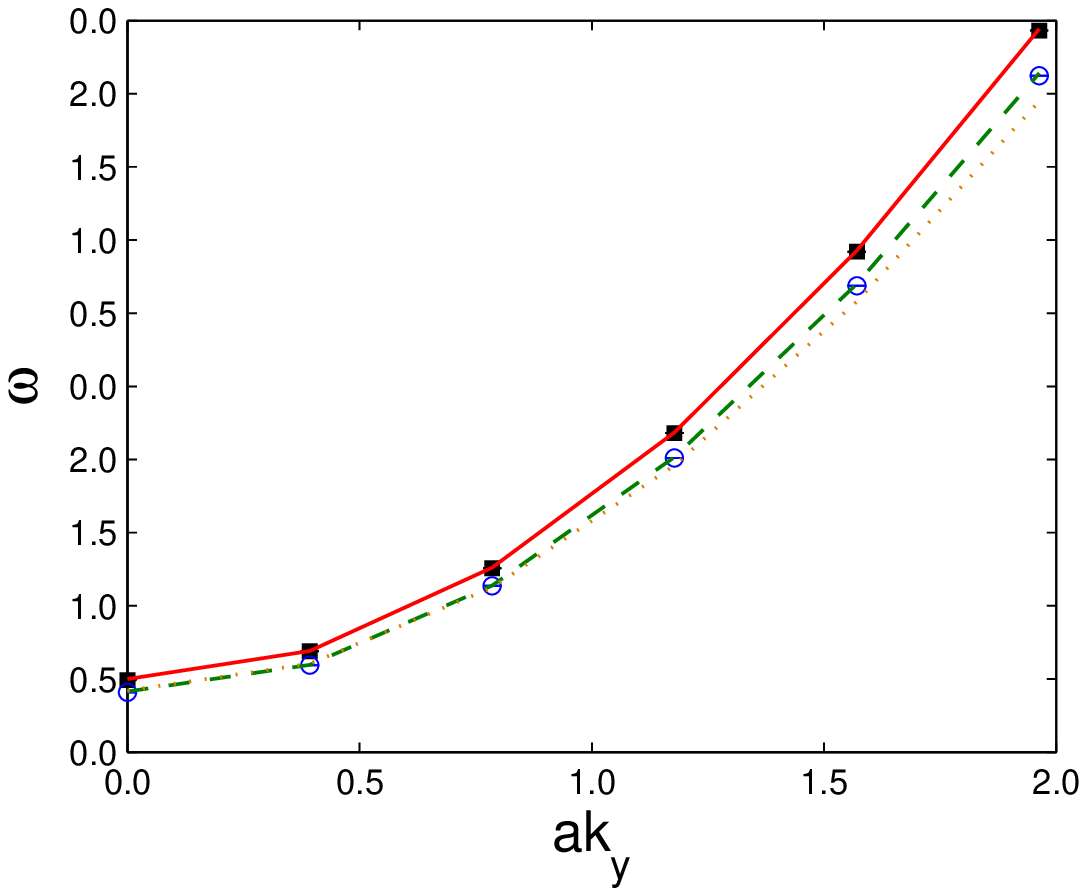}
\caption{Calculated magnon frequencies for wave vectors $\boldsymbol{k}_{i}=(0,k_{y})$, at temperatures $T=0$ and $T=0.2$. For $T=0$, filled squares represent the results of the simulations, while the solid line corresponds to values calculated by using equation (\ref{spekeq}). For $T=0.2$, open circles represent the results of the simulations, the dashed line is calculated by using the variational method and the dotted line using the mean field approach. The parameters are $J_{xx}=J_{yy}=J_{zz}=-1$, $D=0$, $\alpha=0.05$, $K_{x}=-0.1$.}
\label{D0kyT}
\end{figure}

\begin{figure}[H]
\centering
\includegraphics[width=12cm,height=7.5cm]{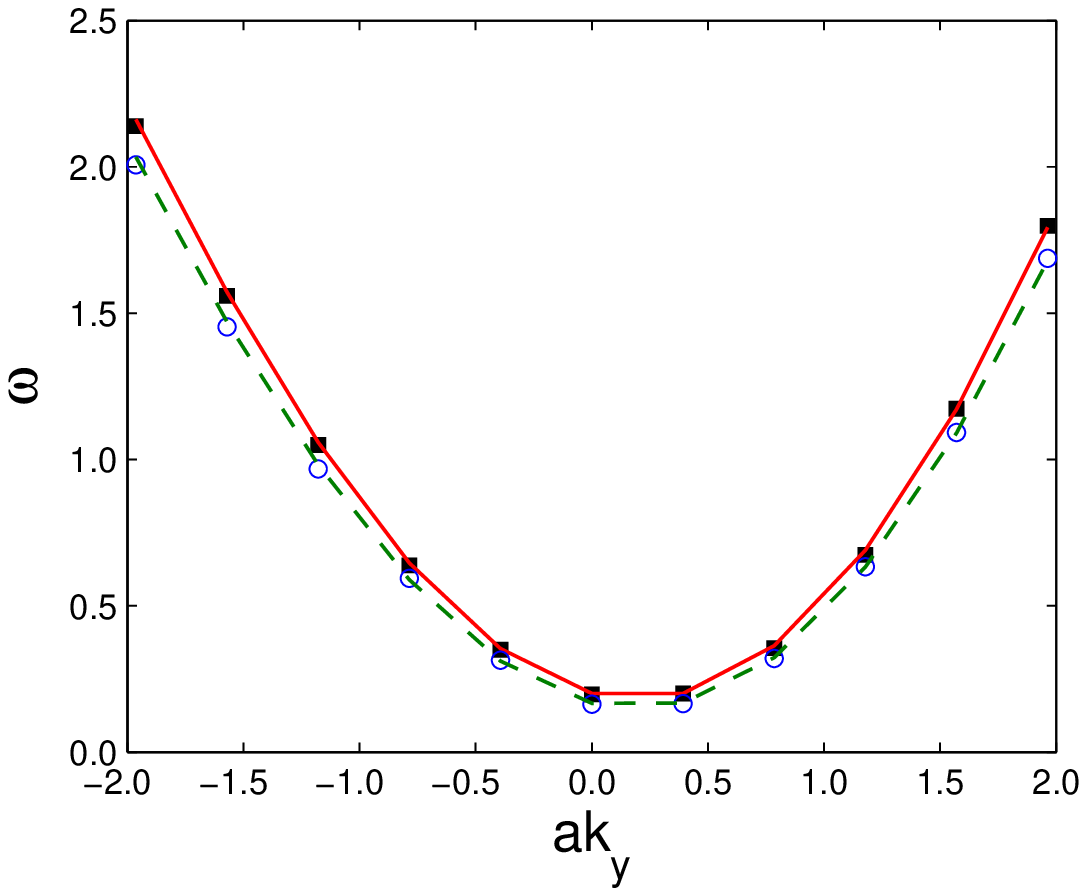}
\caption{Calculated magnon frequencies for wave vectors $\boldsymbol{k}_i=(0,k_y)$, at temperatures $T=0$ and $T=0.2$. For $T=0$, filled squares represent the results of the simulations, while the solid line corresponds to values calculated by using equation (\ref{spekeq}). For $T=0.2$, open circles represent the results of the simulations and the dashed line is calculated by using the variational method. The parameters are $J_{xx}=J_{yy}=J_{zz}=-1$, $D=0.1$, $\alpha=0.05$, $K_x=-0.1$.}
\label{D01kyT}
\end{figure}

Figures \ref{D0kyT} and \ref{D01kyT} show the results for finite wave vectors. At $T=0$, both theoretical methods give the same magnon spectrum apart from a factor of $(1+\alpha^{2})^{-1}$, and this is in relatively good agreement with the simulations. At $T=0.2$, the variational method again recovers the results of the simulations with a high accuracy. The mean field approach reproduces qualitatively well the decrease of the magnon dispersion as compared to $T=0$, but overestimates the magnon frequencies at low wave numbers and underestimates them at high wave numbers. According to the results depicted in figure \ref{D01kyT} the variational method is suitable for describing the magnon spectrum at finite temperatures also in the presence of Dzyaloshinsky-Moriya interactions.

The present results on the temperature dependence of the magnon dispersion make possible to quantitatively revise the spin-wave spectra of Fe/W(110) obtained from ab inito calculations\cite{Costa2010,Udvardi} as compared to the experiments performed at $T=120$ K \cite{Prokop,Zakeri}. Recalling the Curie temperature $T_C=223$ K, this implies a ratio of $T/T_{C} \simeq 0.54$. As mentioned earlier, Monte Carlo simulations resulted in $T_C \simeq 0.7 \mid \!\! J \!\! \mid$ for the present model, thus the temperature of the experiment corresponds to $T\simeq 0.38$ on our temperature scale. From figure \ref{D0ky01Txc} one can then easily extrapolate a value of $\omega \simeq 0.2$ at this temperature, which means a relative decrease of 0.7 for the magnon frequency. Since the energy scale of ab initio magnon spectra corresponding to $T=0$ was typically twice as large as that in the experiment, we can conclude that a large part -- at least, 60\% -- of the difference can be accounted for by the direct effect of finite temperature transversal spin fluctuations on the magnon spectrum. The rest of the difference might be attributed to changes of the spin-Hamiltonian parameters, mainly due to longitudinal spin-fluctuations \cite{Bergman}.

\begin{figure}[H]
\centering
\includegraphics[width=12cm,height=7.5cm]{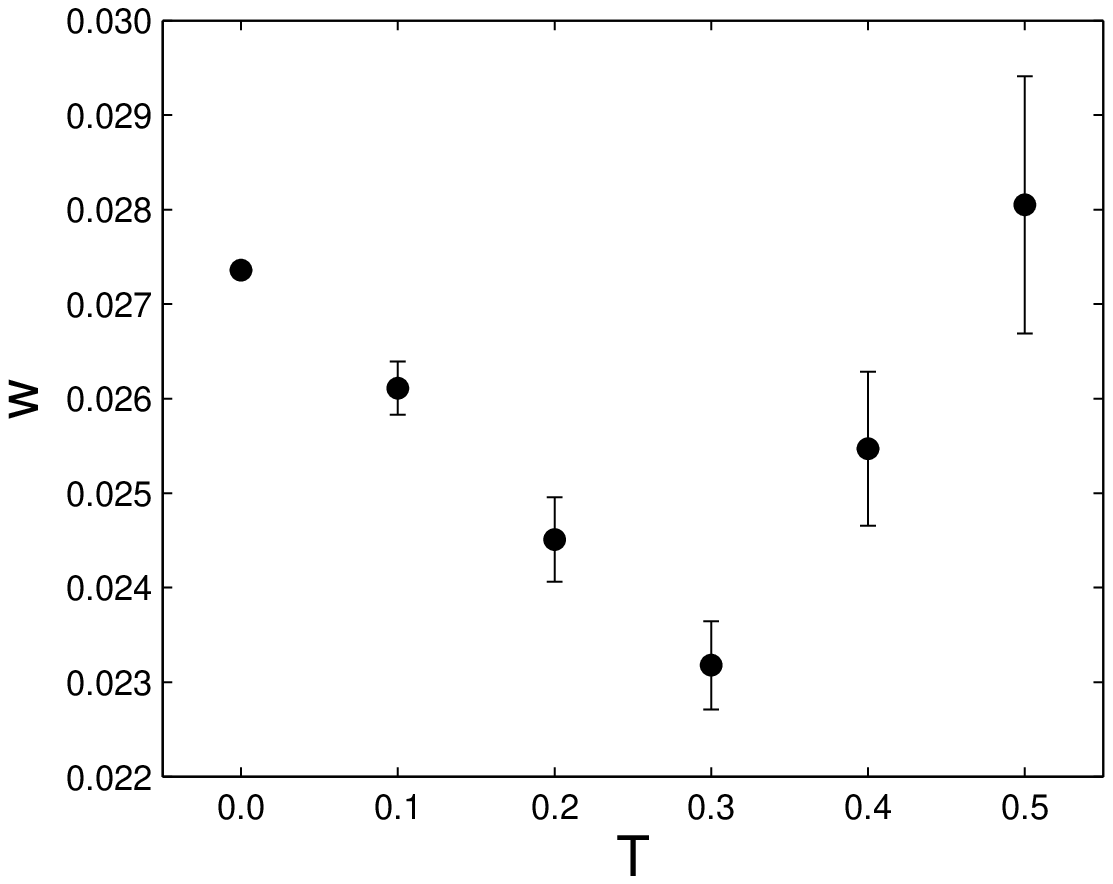}
\caption{The half-width of the resonance curve at the wave vector $(k_{x},k_y)=(0,\frac{2\pi}{16a})$, between temperatures $T=0$ and $T=0.5$. The errorbars correspond to the error of the fitting parameters. The parameters $J_{xx}=J_{yy}=J_{zz}=-1$, $D=0$, $K_x=-0.1$ and $\alpha=0.05$ were used for the simulations. The half-width first decreases with increasing temperature due to the decreasing magnon energy, see equation (\ref{curve}) and figure \ref{D0ky01Txc}, and then starts to increase due to magnon-magnon scattering.}
\label{D0ky01Tw}
\end{figure}

Finally, the half-width $w$ of the resonance curves is investigated as a function of the temperature. In figure \ref{D0ky01Tw} two distinct regions are apparent: the half-width decreases almost linearly up to $T=0.3$, then it rapidly increases. To explain this behavior, we recall expression (\ref{curve}) indicating that, at $T=0$, $w \simeq 2\alpha\omega_{0}$, where $\omega_{0}$ is the resonance frequency. Supposing a similar relationship at low temperatures, the decrease of $\omega_{\boldsymbol{k}}(T)$ with increasing $T$, see figure \ref{D0ky01Txc}, implies a decrease of the half-width, with an almost strict proportionality with $T$. This effect is present in the linear response theory based on the stochastic Landau-Lifshitz-Gilbert equations\cite{Raikher}, but only in the non-interacting case. Within our mean field treatment for an interacting spin-model, see(\ref{szsol}), the half-width always increases with increasing temperature, which is reflected in the simulations only at higher temperatures. Note that in a quantum theoretical description\cite{Costa Filho, Lutsev}, this widening corresponds to the decrease of the magnon lifetime due to increased magnon-magnon scattering.

\section{Summary and conclusions}

A model Hamiltonian of a ferromagnetic thin film was examined using the quasiclassical stochastic Landau-Lifshitz-Gilbert equations (\ref{LLG}). The model included exchange interaction, Dzyaloshinsky-Moriya interaction as well as on-site and two-site anisotropy terms. It was found that at zero temperature and close to the ground state, the linear response of the system to periodic external excitation can be described by a resonance curve (\ref{curve}), and the locations of the peaks on this curve can be interpreted as magnon energies. The expression (\ref{spekeq}) for the magnon dispersion relation at zero temperature unambiguously confirms that the anisotropy terms induce a gap, while the Dzyaloshinsky-Moriya interaction leads to an asymmetry in the spectrum.

The effect of lattice defects was also investigated for the zero wave vector excitation in an approximation of the dynamical equations. It was found that the presence of defects leads to the softening of the magnon energy if two-site anisotropy or Dzyaloshinsky-Moriya interaction is present in the system, but the lineshape does not change if only on-site anisotropy is considered. Since the chiral interaction has no effect on the magnon energy in the $\Gamma$ point in a perfect lattice, this decrease of the magnon energy must be a consequence of magnon scattering at lattice defects.

Two models were discussed for finite temperature effects. The mean field result (\ref{szsol}) was based on the solution of the moment equations calculated from the dynamical equations. For simplicity, we solved the equations for the case when the Dzyaloshinsky-Moriya interaction was absent. The solution indicated that the magnon energy decreases with increasing temperature, while the linewidth of the resonance curve increases. The variational method (\ref{scons1})-(\ref{scons2}) was based on a quantumtheoretical treatment: here we included the Dzyaloshinsky-Moriya interaction, but obtained information only for the magnon energies which, in correspondence with the mean field result, also indicated a decreasing behaviour with increasing temperature.

The theoretical calculations were then compared to numerical spin dynamics simulations, where the linear response of the system was calculated as a function of the frequency of an external excitation field perpendicular to the magnetization. The dynamic structure factor was also calculated, providing information about the magnon spectrum in the whole Brillouin zone. The simulations were in generally good agreement with the theoretical results, although it was found that the lineshape of the linear response curve becomes asymmetric by decreasing the damping parameter. For the finite temperature magnon energies, both theoretical methods and the simulations provided consistent results. The mean field approach showed increasing deviations at higher temperatures and higher wave vectors as compared to the other method. The results were used to provide a quantitative correction to the ab inito spin wave spectrum on Fe/W($110$) related to finite temperature effects that made possible a better comparison with available experiments. It was also found that the resonance linewidth in the simulations first decreases with increasing temperature, then rapidly increases, while within the mean field approximation the decreasing part is absent.

In conclusion, the results of the theoretical methods presented here were in good agreement with numerical simulations. It is worth to generalize them to other lattice types, including three dimensional structures, to other types of interactions, or even to antiferromagnets. Finally, it is worth to mention that strong effects of point-like defects on the magnon lifetime in noncollinear antiferromagnets have very recently been demonstrated by Brenig {\it et al.}\cite{Brenig}.

\ack
The authors thank Ulrich Nowak and Denise Hinzke for enlightening discussions. Financial support was provided by the Hungarian National Research Foundation (under contracts OTKA  77771 and 84078).

\appendix
\setcounter{section}{1}

\section*{Appendix}

\subsection{The linearized equations of motion in case of lattice defects\label{lattdefapp}}

As noted in section \ref{lattdef}, one has to consider the angle variables at the four nearest-neighbour and the two next-nearest-neighbour sites sites of a vacancy separately from the others, since when one writes down equations (\ref{eq1})-(\ref{eq2}) for these lattice points, some terms will be missing. Lattice points further away from the vacancy will be described by an ''average'' angle variable, since all of their nearest-neighbour and next-nearest-neighbour lattice sites contain a spin. Of course these points are not exactly equivalent, since a lattice point that is two lattice vectors far from the defect has a common neighbour with the vacancy, while this is not true for lattice points further away. This difference is only noticeable in higher order calculations.

\begin{figure}[H]
\centering
\includegraphics[width=7.5cm,height=12cm]{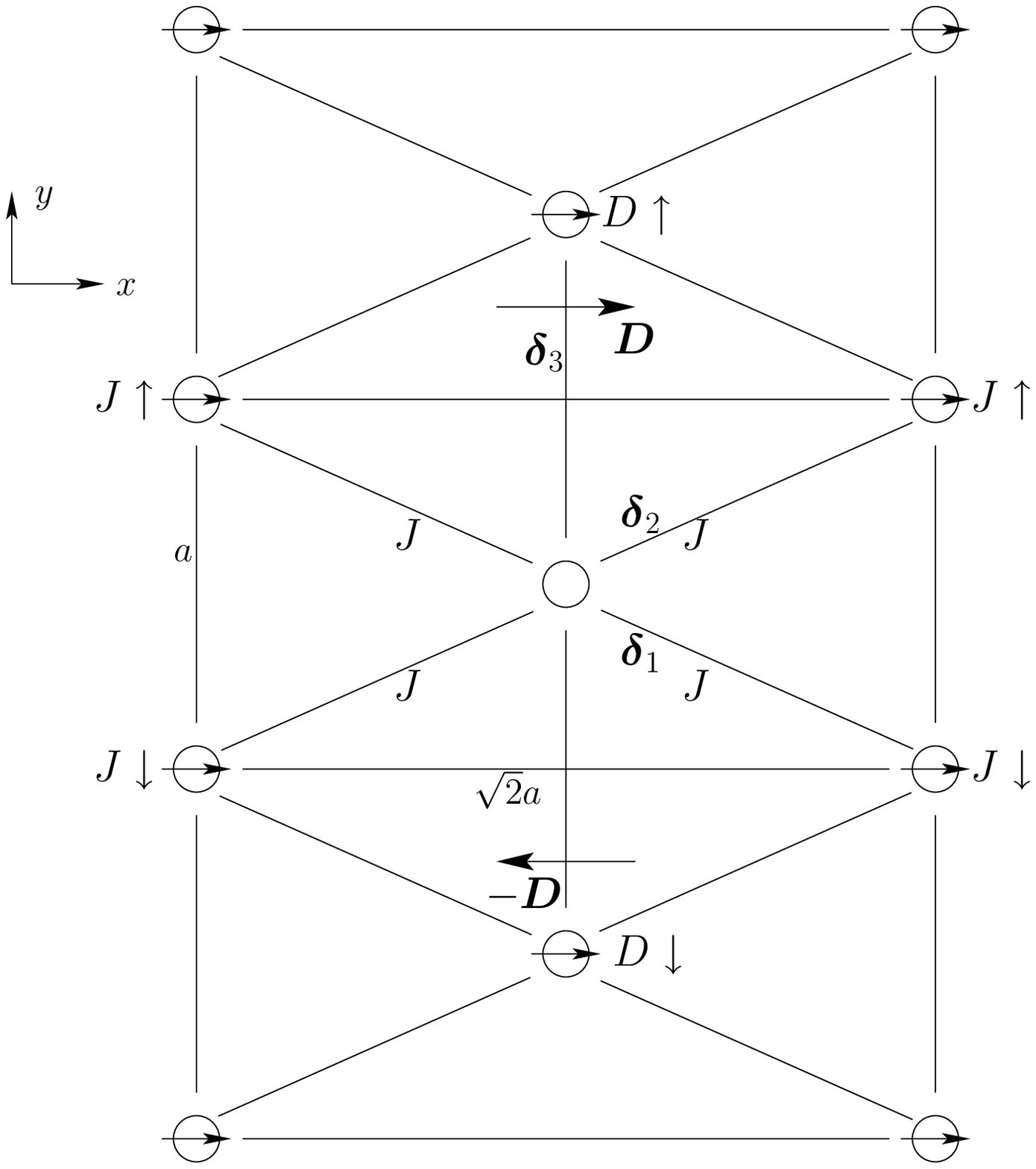}
\caption{Similar to figure \ref{szomszed} with the notation used in \ref{lattdefapp} to describe the
spins at the nearest neighbours and next-nearest neighbours of a lattice vacancy, $J\uparrow$, $J\downarrow$, $D\uparrow$, and $D\downarrow$.}
\label{nearvac}
\end{figure}

For this ''average'' angle variable, one has to sum up the linearized equations (\ref{eq1})-(\ref{eq2}) for all lattice points, and divide them with the number of lattice points. Here the effect of the vacancy is that a term must be subtracted from the original equations, corresponding to the nearest-neighbour and next-nearest-neighbour sites of the defect. Since in a large enough lattice the effect of a single defect is hardly noticeable, multiple vacancies can be considered by multiplying this subtracted term by their number, this is a relatively good approximation if the vacancies are still far from each other. This means that if in a lattice with $n$ points, $m$ spins are set to zero, the correct normalization for the ''average'' variable is $\hat{\beta}_{\alpha}=\frac{1}{n-m}\sum_{i}\beta_{\alpha, i}$ for $\alpha=1, 2$. Using the notations in figure \ref{nearvac}, it can be seen that out of the six lattice points next to the vacancies, only four of them lead to different equations. Therefore, the following ten linear differential equations are to be solved simultaneously,
\begin{eqnarray}
(1+\alpha^2)\frac{\rmd\beta_{2J\uparrow}}{\rmd t}= - 3J_{xx}\beta_{1J\uparrow} + 2J_{zz}\hat{\beta}_{1} + J_{zz}\beta_{1D\uparrow} + D\hat{\beta}_{2} - D\beta_{2J\downarrow} \nonumber
\\
\qquad- 2K_{x}\beta_{1J\uparrow} - \alpha\bigg(- 3J_{xx}\beta_{2J\uparrow} + 2J_{yy}\hat{\beta}_{2} + J_{yy}\beta_{2D\uparrow} - D\hat{\beta}_{1} \nonumber
\\
\qquad+ D\beta_{1J\downarrow} - 2K_{x}\beta_{2J\uparrow}\bigg) + B_{z}\cos{\omega t} \, , \label{tol}
\\
(1+\alpha^2)\frac{\rmd\beta_{1J\uparrow}}{\rmd t}= - \bigg(- 3J_{xx}\beta_{2J\uparrow} + 2J_{yy}\hat{\beta}_{2} + J_{yy}\beta_{2D\uparrow} - D\hat{\beta}_{1} + D\beta_{1J\downarrow} \nonumber
\\
\qquad- 2K_{x}\beta_{2J\uparrow}\bigg) - \alpha\bigg(- 3J_{xx}\beta_{1J\uparrow} + 2J_{zz}\hat{\beta}_{1} + J_{zz}\beta_{1D\uparrow} + D\hat{\beta}_{2} \nonumber
\\
\qquad- D\beta_{2J\downarrow} - 2K_{x}\beta_{1J\uparrow}\bigg) - \alpha B_{z}\cos{\omega t} \, ,
\end{eqnarray}
\begin{eqnarray}
(1+\alpha^2)\frac{\rmd\beta_{2J\downarrow}}{\rmd t}= - 3J_{xx}\beta_{1J\downarrow} + 2J_{zz}\hat{\beta}_{1} + J_{zz}\beta_{1D\downarrow} - D\hat{\beta}_{2} + D\beta_{2J\uparrow} \nonumber
\\
\qquad- 2K_{x}\beta_{1J\downarrow} - \alpha\bigg(- 3J_{xx}\beta_{2J\downarrow} + 2J_{yy}\hat{\beta}_{2} + J_{yy}\beta_{2D\downarrow} + D\hat{\beta}_{1} \nonumber
\\
\qquad- D\beta_{1J\uparrow} - 2K_{x}\beta_{2J\downarrow}\bigg) + B_{z}\cos{\omega t} \, ,
\\
(1+\alpha^2)\frac{\rmd\beta_{1J\downarrow}}{\rmd t}= - \bigg(- 3J_{xx}\beta_{2J\downarrow} + 2J_{yy}\hat{\beta}_{2} + J_{yy}\beta_{2D\downarrow} + D\hat{\beta}_{1} - D\beta_{1J\uparrow} \nonumber
\\
\qquad- 2K_{x}\beta_{2J\downarrow}\bigg) - \alpha\bigg(- 3J_{xx}\beta_{1J\downarrow} + 2J_{zz}\hat{\beta}_{1} + J_{zz}\beta_{1D\downarrow} - D\hat{\beta}_{2} \nonumber
\\
\qquad+ D\beta_{2J\uparrow} - 2K_{x}\beta_{1J\downarrow}\bigg) - \alpha B_{z}\cos{\omega t} \, ,
\end{eqnarray}
\begin{eqnarray}
(1+\alpha^2)\frac{\rmd\beta_{2D\uparrow}}{\rmd t}= - 4J_{xx}\beta_{1D\uparrow} + 2J_{zz}\hat{\beta}_{1} + 2J_{zz}\beta_{1J\uparrow} + D\hat{\beta}_{2} \nonumber
\\
\qquad- 2K_{x}\beta_{1D\uparrow} - \alpha\bigg(- 4J_{xx}\beta_{2D\uparrow} + 2J_{yy}\hat{\beta}_{2} + 2J_{yy}\beta_{2J\uparrow} \nonumber
\\
\qquad- D\hat{\beta}_{1} - 2K_{x}\beta_{2D\uparrow}\bigg) + B_{z}\cos{\omega t} \, ,
\\
(1+\alpha^2)\frac{\rmd\beta_{1D\uparrow}}{\rmd t}= - \bigg(- 4J_{xx}\beta_{2D\uparrow} + 2J_{yy}\hat{\beta}_{2} + 2J_{yy}\beta_{2J\uparrow} - D\hat{\beta}_{1} \nonumber
\\
\qquad- 2K_{x}\beta_{2D\uparrow}\bigg) - \alpha\bigg(- 4J_{xx}\beta_{1D\uparrow} + 2J_{zz}\hat{\beta}_{1} + 2J_{zz}\beta_{1J\uparrow} \nonumber
\\
\qquad+ D\hat{\beta}_{2} - 2K_{x}\beta_{1D\uparrow}\bigg) - \alpha B_{z}\cos{\omega t} \, ,
\end{eqnarray}
\begin{eqnarray}
(1+\alpha^2)\frac{\rmd\beta_{2D\downarrow}}{\rmd t}= - 4J_{xx}\beta_{1D\downarrow} + 2J_{zz}\hat{\beta}_{1} + 2J_{zz}\beta_{1J\downarrow} - D\hat{\beta}_{2} \nonumber
\\
\qquad- 2K_{x}\beta_{1D\downarrow} - \alpha\bigg(- 4J_{xx}\beta_{2D\downarrow} + 2J_{yy}\hat{\beta}_{2} + 2J_{yy}\beta_{2J\downarrow} \nonumber
\\
\qquad+ D\hat{\beta}_{1} - 2K_{x}\beta_{2D\downarrow}\bigg) + B_{z}\cos{\omega t} \, ,
\\
(1+\alpha^2)\frac{\rmd\beta_{1D\downarrow}}{\rmd t}= - \bigg(- 4J_{xx}\beta_{2D\downarrow} + 2J_{yy}\hat{\beta}_{2} + 2J_{yy}\beta_{2J\downarrow} + D\hat{\beta}_{1} \nonumber
\\
\qquad- 2K_{x}\beta_{2D\downarrow}\bigg) - \alpha\bigg(- 4J_{xx}\beta_{1D\downarrow} + 2J_{zz}\hat{\beta}_{1} + 2J_{zz}\beta_{1J\downarrow} \nonumber
\\
\qquad- D\hat{\beta}_{2} - 2K_{x}\beta_{1D\downarrow}\bigg) - \alpha B_{z}\cos{\omega t} \, ,
\end{eqnarray}
\begin{eqnarray}
(1+\alpha^2)\frac{\rmd\hat{\beta}_{2}}{\rmd t}=\bigg[-4(J_{xx}-J_{zz}) - 2K_{x} + B_{x}\bigg] \hat{\beta}_{1} \nonumber
\\
\qquad- \alpha\bigg[-4(J_{xx}-J_{yy}) - 2K_{x}\bigg] \hat{\beta}_{2} - \frac{m}{n-m}\Bigg\{-2(J_{xx}-J_{zz}) \nonumber
\\
\qquad\times(\beta_{1J\uparrow}+\beta_{1J\downarrow}) + D(\beta_{2D\uparrow}-\beta_{2D\downarrow}) - \alpha\bigg[-2(J_{xx}-J_{yy}) \nonumber
\\
\qquad\times(\beta_{2J\uparrow}+\beta_{2J\downarrow}) - D(\beta_{1D\uparrow}-\beta_{1D\downarrow})\bigg]\Bigg\} + B_{z}\cos{\omega t} \, ,\label{ig1}
\\
(1+\alpha^2)\frac{\rmd\hat{\beta}_{1}}{\rmd t}=- \bigg[-4(J_{xx}-J_{yy}) - 2K_{x}\bigg) \hat{\beta}_{2} \nonumber
\\
\qquad- \alpha\bigg(-4(J_{xx}-J_{zz}) - 2K_{x}\bigg] \hat{\beta}_{1} - \frac{m}{n-m}\Bigg\{- \bigg[-2(J_{xx}-J_{yy}) \nonumber
\\
\qquad\times(\beta_{2J\uparrow}+\beta_{2J\downarrow}) - D(\beta_{1D\uparrow}-\beta_{1D\downarrow})\bigg] - \alpha\bigg[-2(J_{xx}-J_{zz}) \nonumber
\\
\qquad\times(\beta_{1J\uparrow}+\beta_{1J\downarrow}) + D(\beta_{2D\uparrow}-\beta_{2D\downarrow})\bigg]\Bigg\} - \alpha B_{z}\cos{\omega t} \, . \label{ig}
\end{eqnarray}

Similarly to expression (\ref{S}), the linear response of the system at $\boldsymbol{k}_{i}=\boldsymbol{0}$ is now defined as
\begin{eqnarray}
S(\omega)=\langle \hat{\beta}_{2}^{2} \rangle + \langle \hat{\beta}_{1}^{2} \rangle - \langle \hat{\beta}_{2} \rangle^{2} - \langle \hat{\beta}_{1} \rangle^{2} \, .
\end{eqnarray}

Equations (\ref{tol})-(\ref{ig}) are equivalent of a system of linear algebraic equations through Fourier transformation in time, which can be easily solved numerically. It is important to note that in the absence of Dzyaloshinsky-Moriya interaction ($D=0$) and two-site anisotropy ($J_{xx}=J_{yy}=J_{zz}$) equations (\ref{ig1})-(\ref{ig}) are uncoupled from the first eight ones, and with the applied normalization they do not depend on the number of atoms in the lattice. Therefore it can be concluded that  $S(\omega)$ is not affected by lattice defects if only isotropic exchange interactions and on-site anisotropy are present in the system.

\end{document}